\documentclass[article, nojss]{jss}

\usepackage{amsmath}
\usepackage{amssymb}
\usepackage{bm}
\usepackage{textcomp}
\usepackage{tablefootnote}

\newcommand{\mupar}{\mu}
\newcommand{\phipar}{\phi}
\newcommand{\sigmapar}{{\sigma_\eta}}
\newcommand{\sigmapartwo}{{\sigma_\eta^2}}
\newcommand{\sigmaparextrasub}[1]{\sigma_{\eta,{#1}}}
\newcommand{\allpara}{\bm{\theta}}

\newcommand{\hsv}{h}
\newcommand{\hsvm}{\hb}
\newcommand{\y}{y}
\newcommand{\unobs}{\bm{\kappa}}
\newcommand{\sigmaeps}{{\sigma_\epsilon}}
\newcommand{\sigmaepsextrasub}[1]{\sigma_{\epsilon,{#1}}}
\newcommand{\sigmaepsextrasuper}[1]{\sigma^{#1}_\epsilon}

\newcommand{\ML}{M\!L}
\newcommand{\PL}{P\!L}

\newcommand{\Xb}{\bm{X}}
\newcommand{\yb}{\bm{y}}
\newcommand{\betab}{\bm{\beta}}
\newcommand{\Ib}{\bm{I}}
\newcommand{\bb}[1]{\bm{b}_{#1}}
\newcommand{\Bb}[1]{\bm{B}_{#1}}
\newcommand{\bpostb}{\bm{b}_T}
\newcommand{\Bpostb}{\bm{B}_T}
\newcommand{\Sigmab}{\bm{\Sigma}}
\newcommand{\hb}{\bm{h}}
\newcommand{\pb}{\bm{p}}

\newcommand{\XtX}{\Xb^\top \Xb}

\newcommand{\Normal}[1]{\mathcal{N}\!\left(#1\right)}

\newcommand{\Gammainv}[1]{\mathcal{G}^{-1}\!\left(#1\right)}
\newcommand{\diag}[1]{\text{diag}({#1})}
\newcommand{\Betafun}[1]{B (#1)}
\newcommand{\Betadis}[1]{\mathcal{B}\left(#1\right)}
\newcommand{\Gammad}[1]{ \mathcal{G}\left(#1\right)}

\newcommand{\e}{e}
\newcommand{\dif}{\mathrm{d}}

\newtheorem{alg}{Algorithm}

\graphicspath{{extrafig/}}

\author{Gregor Kastner\\WU Vienna University of Economics and Business}
\title{Dealing with Stochastic Volatility in Time Series Using the \proglang{R} Package \pkg{stochvol}}
\Plaintitle{Dealing with Stochastic Volatility in Time Series Using the R Package stochvol} 

\Plainauthor{Gregor Kastner} 

\Abstract{
The \proglang{R} package \pkg{stochvol} provides a fully Bayesian implementation of heteroskedasticity modeling within the framework of stochastic volatility.
It utilizes Markov chain Monte Carlo (MCMC) samplers to conduct inference by obtaining draws from the posterior distribution of parameters and latent variables which can then be used for predicting future volatilities. The package can straightforwardly be employed as a stand-alone tool; moreover, it allows for easy incorporation into other MCMC samplers. The main focus of this paper is to show the functionality of \pkg{stochvol}. In addition, it provides a brief mathematical description of the model, an overview of the sampling schemes used, and several illustrative examples using exchange rate data.
}

\Address{
  Gregor Kastner\\
  Institute for Statistics and Mathematics\\
  Department of Finance, Accounting and Statistics\\
  WU Vienna University of Economics and Business\\
  Welthandelsplatz 1, Building D4, Level 3\\
  1020 Vienna, Austria\\
  Telephone: +43/1/31336-5593\\
  Fax: +43/1/31336-90-5593\\
  E-mail: \email{gregor.kastner@wu.ac.at}\\
  URL: \url{http://statmath.wu.ac.at/~kastner/}
}

\begin{document}

\shortcites{r:mvt}

\section*{Preface}
This vignette corresponds to an article of the same name which is published in the Journal of Statistical Software. At the time of this writing, the vignette and the published article essentially coincide (in some trace plots within this vignette, only every 10th value is shown to keep \pkg{stochvol} below the 5MB mark). The version at hand might receive minor updates as time goes by. To cite, please use \cite{kas:dea}. Further information about citing \pkg{stochvol} can be obtained in \proglang{R} by installing the package, e.g., through \code{install.packages("stochvol")}, and calling \code{citation("stochvol")}.

\section{Introduction}
Returns -- in particular financial returns -- are commonly analyzed by estimating and predicting potentially time-varying volatilities. This focus has a long history, dating back at least to \cite{mar:por} who investigates portfolio construction with optimal expected return-variance trade-off. In his article, he proposes rolling-window-type estimates for the instantaneous volatilities, but already then recognizes the potential for ``better methods, which take into account more information''.

One approach is to model the evolution of volatility deterministically, i.e., through the (G)ARCH class of models. After the groundbreaking papers of \cite{eng:aut} and \cite{bol:gen}, these models have been generalized in numerous ways and applied to a vast amount of real-world problems. As an alternative, \cite{tay:fin} proposes in his seminal work to model the volatility probabilistically, i.e., through a state-space model where the logarithm of the squared volatilities -- the latent states -- follow an autoregressive process of order one. Over time, this specification became known as the \emph{stochastic volatility (SV)} model. Even though several papers \citep[e.g.,][]{jac-etal:bayJBES, ghy-etal:sto, kim-etal:sto} provide early evidence in favor of using SV, these models have found comparably little use in applied work. This obvious discrepancy is discussed in \cite{bos:rel} who points out two reasons: the variety (and potential incompatibility) of estimation methods for SV models -- whereas the many variants of the GARCH model have basically a single estimation method -- and the lack of standard software packages implementing these methods.
 

 In \cite{kas-fru:anc}, the former issue is thoroughly investigated and an efficient MCMC estimation scheme
is proposed. The paper at hand and the corresponding package \pkg{stochvol} \citep{r:sto} for \proglang{R} \citep{r:r} are crafted to cope with the latter problem: the apparent lack of ready-to-use software packages for efficiently estimating SV models.

\section{Model specification and estimation}

We begin by briefly introducing the model and specifying the notation used in the remainder of the paper. Furthermore, an overview of Bayesian parameter estimation via Markov chain Monte Carlo (MCMC) methods is given.

\subsection{The SV model}

Let $\yb = (y_1, y_2, \dots, y_n)^\top$ be a vector of returns with mean zero. The intrinsic feature of the SV model is that each observation $y_t$ is assumed to have its ``own'' contemporaneous variance $\e^{\hsv_t}$, thus relaxing the usual assumption of homoskedasticity. In order to make the estimation of such a model feasible, 
this variance is not allowed to vary unrestrictedly with time. Rather, its logarithm is assumed to follow an autoregressive process of order one. Note that this feature is fundamentally different to GARCH-type models where the time-varying volatility is assumed to follow a deterministic instead of a stochastic evolution.

The SV model can thus be conveniently expressed in hierarchical form. In its centered parameterization, it is given through
\begin{eqnarray}  
 \label{c1}
 y_{t}|\hsv_t &\sim&  \Normal{0, \exp{\hsv_{t}}},
 \\
 \label{c2}
 \hsv_{t}|\hsv_{t-1},\mupar, \phipar, \sigmapar&\sim& \Normal{\mupar +  \phipar (\hsv_{t-1}- \mupar ),  \sigmapartwo},\\
 \label{c3}
\hsv_0|\mu,\phi,\sigmapar &\sim& \Normal{\mu,\sigmapartwo/(1-\phi^2)},
\end{eqnarray}

where $\Normal{\mu, \sigmapartwo}$ denotes the normal distribution with mean $\mu$ and variance $\sigmapartwo$. We refer to $\allpara = (\mupar, \phipar, \sigmapar)^\top$ as the vector of \emph{parameters}: the \emph{level} of log-variance $\mupar$, the \emph{persistence} of log-variance $\phipar$, and the \emph{volatility} of log-variance $\sigmapar$. The process $\hsvm=(\hsv_{0}, \hsv_{1}, \ldots,\hsv_{n})$ appearing in Equation~\ref{c2} and Equation~\ref{c3} is unobserved and usually interpreted as the latent time-varying \emph{volatility process} (more precisely, the log-variance process). Note that the initial state $\hsv_0$ appearing in Equation~\ref{c3} is distributed according to the stationary distribution of the autoregressive process of order one.

\subsection{Prior distribution}
\label{priors}

To complete the model setup, a prior distribution for the parameter vector $\allpara$ needs to be specified. Following \cite{kim-etal:sto}, we choose independent components for each parameter, i.e., $p(\allpara) = p(\mupar)p(\phipar)p(\sigmapar)$.

The level $\mupar \in \mathbb{R}$ is equipped with the usual normal prior $\mupar \sim \Normal{b_{\mupar}, B_{\mupar}}$. In practical applications, this prior is usually chosen to be rather uninformative, e.g., through setting $b_\mupar = 0$ and $B_\mupar \geq 100$ for daily log returns. Our experience with empirical data is that the exact choice is usually not very influential; see also Section~\ref{priorconfig}.

For the persistence parameter $\phipar \in (-1,1)$, we choose $(\phi+1)/2 \sim \Betadis{a_0, b_0}$, implying 
\begin{eqnarray}
 \label{beta_transformed}
 p(\phi) = \frac{1}{2\Betafun{a_0,b_0}}\left ( \frac{1+\phi}{2}\right ) ^{a_0-1}\left ( \frac{1-\phi}{2}\right ) ^{b_0-1},
\end{eqnarray}
where $a_0$ and $b_0$ are positive hyperparameters and $\Betafun{x,y} = \int_0^1t^{x-1}(1-t)^{y-1}\,dt$ denotes the beta function.
Clearly, the support of this distribution is the interval $(-1, 1)$; thus, stationarity of the autoregressive volatility process is guaranteed. Its expected value and variance are given through the expressions
\begin{eqnarray*}
 E(\phipar) &=& \frac{2a_0}{a_0+b_0}-1,\\
 V(\phipar) &=& \frac{4a_0b_0}{(a_0+b_0)^2(a_0+b_0+1)}.
\end{eqnarray*}
This obviously implies that the prior expectation of $\phipar$ depends only on the ratio $a_0:b_0$. It is greater than zero if and only if $a_0 > b_0$ and smaller than zero if and only if $a_0 < b_0$. For a fixed ratio $a_0:b_0$, the prior variance decreases with larger values of $a_0$ and $b_0$. The uniform distribution on $(-1,1)$ arises as a special case when $a_0=b_0=1$. For financial datasets with not too many observations (i.e., $n \lesssim 1000$), the choice of the hyperparameters $a_0$ and $b_0$ can be quite influential on the shape of the posterior distribution of $\phipar$. In fact, when the underlying data-generating process is (near-)homoskedastic, the volatility of log-variance $\sigmapar$ is (very close to) zero and thus the likelihood contains little to no information about $\phipar$. Consequently, the posterior distribution of $\phipar$ is (almost) equal to its prior, no matter how many data points are observed. For some discussion about this issue, see, e.g., \cite{kim-etal:sto} who choose $a_0=20$ and $b_0=1.5$, implying a prior mean of $0.86$ with a prior standard deviation of $0.11$ and thus very little mass for nonpositive values of $\phi$.

For the volatility of log-variance $\sigmapar \in \mathbb{R}^+$, we choose $\sigmapartwo \sim B_{\sigmapar}\times \chi^2_1=\Gammad{1/2,1/2B_{\sigmapar}}$. This choice is motivated by \cite{fru-wag:sto} who equivalently stipulate the prior for $\pm \sqrt { \sigmapartwo}$ to follow a centered normal distribution, i.e., $\pm \sqrt { \sigmapartwo} \sim \Normal{0,B_{\sigmapar}}$. As opposed to the more common Inverse-Gamma prior for $\sigmapartwo$, this prior is not conjugate in the usual sampling scheme. However, it does not bound $\sigmapartwo$ away from zero a priori. The choice of the hyperparameter $B_{\sigmapar}$ turns out to be of minor influence in empirical applications as long as it is not set too small.

\subsection{MCMC sampling}

An MCMC algorithm such as the one implemented in the package \pkg{stochvol} provides its user with draws from the posterior distribution of the desired random variables; in our case with the latent log-variances $\hsvm$ and the parameter vector $\allpara$. Because these draws are usually dependent, Bayesian inference via MCMC may require careful design of the algorithm and attentive investigation of the draws obtained.

One key feature of the algorithm used in this package is the joint sampling of all instantaneous volatilities ``all without a loop'' (AWOL), a technique going back at least to \cite{rue:fas} and discussed in more detail in \cite{mcc-etal:sim}. Doing so reduces correlation of the draws significantly and requires auxiliary finite mixture approximation of the errors as in \cite{kim-etal:sto} or \cite{omo-etal:sto}.

In order to avoid the cost of code interpretation within each MCMC iteration, the core computations are implemented in \proglang{C}. Their output is interfaced to \proglang{R} via the \pkg{Rcpp} package \citep{edd-fra:rcp}; there, the convenience functions and the user interface are implemented. This combination allows to make use of the well-established and widely accepted ease-of-use of \proglang{R} and its underlying functional programming paradigm. Moreover, existing frameworks for analyzing MCMC output such as \pkg{coda} \citep{plu-etal:cod} as well as high-level visualization tools can easily be used. Last but not least, users with a basic knowledge of \proglang{R} can use the package with a very low entry cost. Nevertheless, despite all these convenience features, the package profits from a highly optimized machine code generated by a compiler at package build time, thus providing acceptable runtime even for larger datasets.

A novel and crucial feature of the algorithm implemented in \pkg{stochvol} is the usage of a variant of the ``ancillarity-sufficiency interweaving strategy'' (ASIS) which has been brought forward in the general context of state-space models by \cite{yu-men:cen}. ASIS exploits the fact that for certain parameter constellations sampling efficiency improves substantially when considering a non-centered version of a state-space model.
This is commonly referred to as a reparameterization issue and dealt with extensively in scholarly literature; for an early reference see, e.g., \cite{hil-smi:par}.
For the model at hand, a move of this kind can be achieved by transferring the level of log-variance $\mupar$ and/or the volatility of log-variance $\sigmapar$ from the state process (Equations~\ref{c2}~and~\ref{c3}) to the observation process (Equation~\ref{c1}) through a simple reparameterization of $\hsvm$. However, in the case of the SV model, it turns out that no single superior parameterization exists. Rather, for some underlying processes, the standard parameterization yields superior results, while for other processes non-centered versions are better. To overcome this issue, the parameter vector $\allpara$ is sampled twice: once in the centered and once in a noncentered parameterization. This method of ``combining best of different worlds'' allows for efficient inference regardless of the underlying process with one algorithm. For more details about the algorithm and empirical results concerning sampling efficiency, see \cite{kas-fru:anc}.

\section[The stochvol package]{The \pkg{stochvol} package}

The usual stand-alone approach to fitting SV models with \pkg{stochvol} exhibits the following workflow: (1) Prepare the data, (2) specify the prior distributions and configuration parameters, (3) run the sampler, (4) assess the output and display the results. All these steps are described in more detail below, along with a worked example. For a stepwise incorporation of SV effects into other MCMC samplers, see Section~\ref{other}.

\subsection{Preparing the data}
\label{prep}

The core sampling function \code{svsample} expects its input data \code{y} to be a numeric vector of returns without any missing values (\code{NA}s) and throws an error if provided with anything else. In the case that \code{y} contains zeros, a warning is issued and a small offset constant of size \code{sd(y)/10000} is added to the squared returns before doing the auxiliary mixture sampling \citep[cf.][]{omo-etal:sto}. However, we generally recommend to avoid zero returns altogether, e.g., by demeaning them beforehand.

\begin{figure}[!ht]
\begin{center}
 \includegraphics[width=\textwidth]{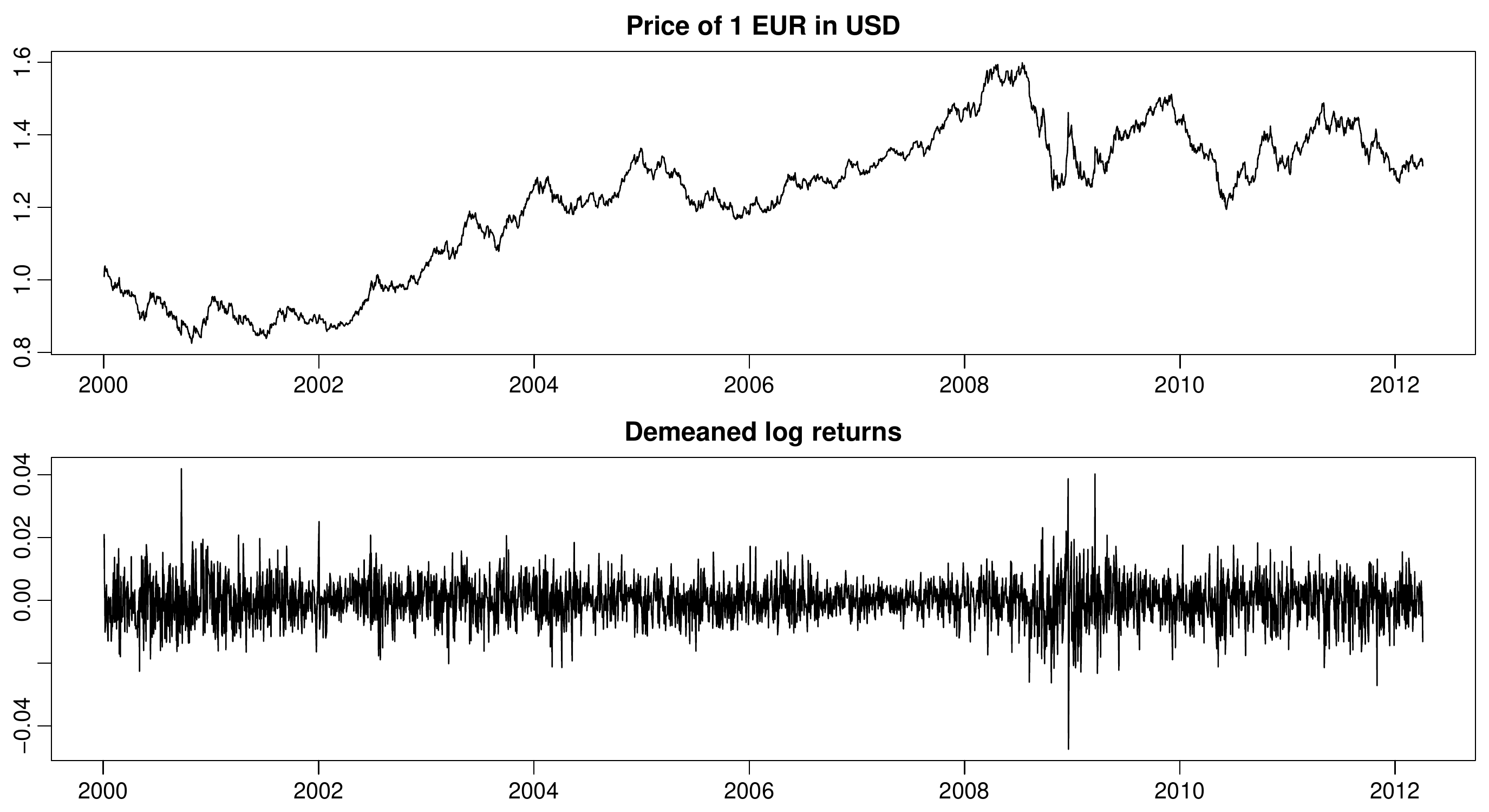}
 \caption{Visualization of EUR-USD exchange rates included in the \pkg{stochvol} package.}
 \label{fig1}
\end{center}
\end{figure}

Below is an illustration of how to prepare data by using the sample dataset \code{exrates}\footnote{The dataset -- obtained from the European Central Bank's Statistical Data Warehouse -- contains the daily bilateral prices of one euro in 23 currencies from January 3, 2000 until April 4, 2012. Conversions to New Turkish lira and Fourth Romanian leu have been incorporated. See \code{?exrates} for more information.} included in the package. Figure~\ref{fig1} provides a visualization of a time series from this dataset.

\begin{Schunk}
\begin{Sinput}
R> set.seed(123)
R> library("stochvol")
R> data("exrates")
R> ret <- logret(exrates$USD, demean = TRUE)
R> par(mfrow = c(2, 1), mar = c(1.9, 1.9, 1.9, 0.5), mgp = c(2, 0.6, 0))
R> plot(exrates$date, exrates$USD, type = "l",
+    main = "Price of 1 EUR in USD")
R> plot(exrates$date[-1], ret, type = "l", main = "Demeaned log returns")
\end{Sinput}
\end{Schunk}

\begin{figure}[!ht]
\begin{center}
 \includegraphics[width=\textwidth]{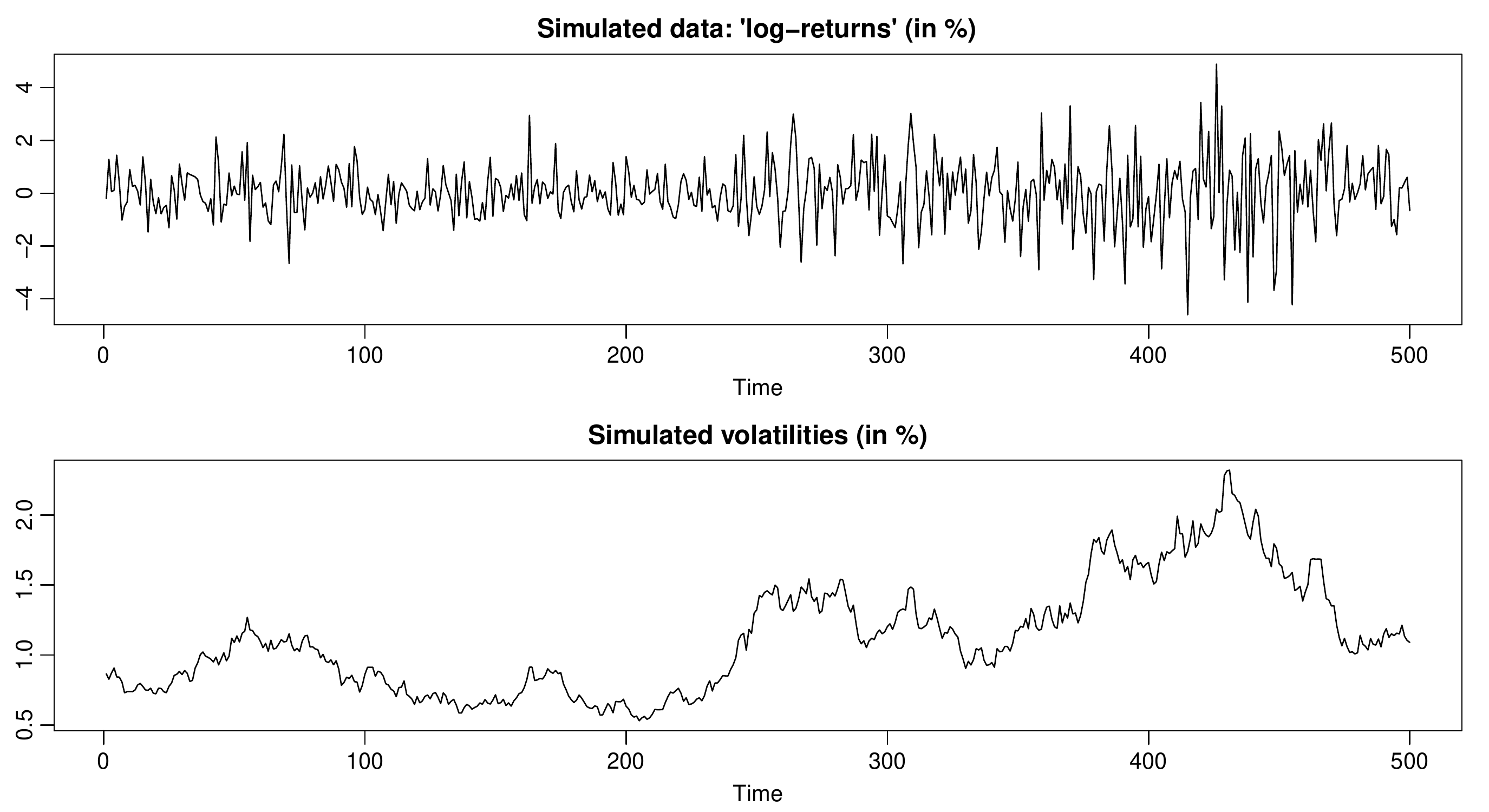}
 \caption{Visualization of a simulated time series as provided by the default \code{plot} method.}
 \label{fig2}
\end{center}
\end{figure}

Additionally to real-world data, \pkg{stochvol} also has a built-in data generator \code{svsim}. This function simply produces realizations of an SV process and returns an object of class \code{svsim} which has its own \code{print}, \code{summary}, and \code{plot} methods. Exemplary code using \code{svsim} is given below and the particular instance of this simulated series is displayed in Figure~\ref{fig2}.

\begin{Schunk}
\begin{Sinput}
R> sim <- svsim(500, mu = -9, phi = 0.99, sigma = 0.1)
R> par(mfrow = c(2, 1))
R> plot(sim)
\end{Sinput}
\end{Schunk}

 \subsection{Specifying prior distributions and configuration parameters}
 \label{priorconfig}

 After preparing the data vector \code{y}, the user needs to specify the prior hyperparameters for the parameter vector $\allpara = (\mupar, \phipar, \sigmapar)^\top$ -- see also Section~\ref{priors} -- and some configuration parameters. The appropriate values are passed to the main sampling function \code{svsample} as arguments which are described below.

 The argument \code{priormu} is a vector of length $2$, containing mean and standard deviation of the normal prior for the level of the log-variance $\mupar$. A common strategy is to choose a vague prior here, e.g., \code{c(0, 100)}, because the likelihood usually carries enough information about this parameter. If one prefers to use (slightly) informative priors, e.g., to avoid outlier draws of $\mupar$, one must pay attention to whether \emph{log returns} or \emph{percentage log returns} are analyzed. Assuming daily data, log returns commonly have an unconditional variance of $0.0001$ or less and thus the level on the log scale $\mupar$ lies around $\log(0.0001) \approx -9$. Percentage log returns, on the other hand, have the $100^2$-fold unconditional variance (around $1$) which implies a level of $\log(1) = 0$. Choices in the literature include \code{c(0, 10)} \citep{jac-etal:bayJE}, \code{c(0, 5)} \citep{yu:on}, \code{c(0, sqrt(10))} \citep{kim-etal:sto, mey-yu:bug} or \code{c(0, 1)} \citep{omo-etal:sto}. Note that most of these choices are quite informative and clearly designed for percentage log returns.

 For specifying the prior hyperparameters for the persistence of log-variance, $\phipar$, the argument \code{priorphi} may be used. It is again a vector of length $2$, containing $a_0$ and $b_0$ specified in Equation~\ref{beta_transformed}. As elaborated in Section~\ref{priors}, these values can possibly be quite influential, thus we advise to choose them carefully and study the effects of different choices. The default is currently given through \code{c(5, 1.5)}, implying a prior mean of $0.54$ and a prior standard deviation of $0.31$.

 The prior variance of log-variance hyperparameter $B_\sigmapar$ may be controlled through \code{priorsigma}. This argument defaults to \code{1} if not provided by the user. As discussed in Section~\ref{priors}, the exact choice of this value is usually not very influential in typical applications. In general, it should not be set too small unless there is a very good reason, e.g., explicit prior knowledge, to do so.

For specifying the size of the burn-in, the parameter \code{burnin} is provided. It is the amount of MCMC iterations that are run but discarded to ensure convergence to the stationary distribution of the chain. The current default value for this parameter is $1000$ which has turned out to suffice in most situations. Nevertheless, the user is encouraged to check convergence carefully; see Section~\ref{check} for more details. The amount of iterations which are run after burn-in can be specified through the parameter \code{draws}, currently defaulting to $10\,000$. Consequently, the sampler is run for a total of \code{burnin + draws} iterations.

 Three thinning parameters are available which all are $1$ if not specified otherwise. The first one, \code{thinpara}, is the denominator in the fraction of parameter draws (i.e., draws of $\allpara$) that are stored. E.g., if \code{thinpara} equals $10$, every $10$th draw is kept. The default parameter thinning value of $1$ means that all draws are saved. The second thinning parameter, \code{thinlatent}, acts in the same way for the latent variables $\hsvm$. The third thinning parameter, \code{thintime}, refers to thinning with respect to the time dimension of the latent volatility.
In the case that \code{thintime} is greater than 1, not all elements of $\hsvm$ are stored, e.g., for \code{thintime} equaling $10$, only the draws of $\hsv_1, \hsv_{11}, \hsv_{21}, \dots$ (and $\hsv_0$) are kept.

Another configuration argument is \code{quiet} which defaults to \code{FALSE}. If set to \code{TRUE}, all output during sampling (progress bar, status messages) is omitted. The arguments \code{startpara} and \code{startlatent} are optional starting values for the parameter vector $\allpara$ and the latent variables $\hsvm$, respectively. All other configuration parameters are summarized in the argument \code{expert}, because it is not very likely that the end-user needs to mess with the defaults.\footnote{Examples of configurations that can be changed with the \code{expert}-argument include the specification of the (baseline) parameterization (either \emph{centered} or \emph{noncentered}) and the possibility to turn off interweaving. Moreover, some algorithmic details such as the number of blocks used for the parameter updates or the possibility of using a random walk Metropolis-Hastings proposal (instead of the default independence proposal) can be found here.} Please refer to the package documentation and \cite{kas-fru:anc} for details.

Any further arguments (\code{\dots}) are forwarded to \code{updatesummary}, controlling the type of summary statistics that are calculated for the posterior draws.

\subsection{Running the sampler}

At the heart of the package \pkg{stochvol} lies the function \code{svsample} which serves as an \proglang{R}-wrapper for the actual sampler coded in \proglang{C}. Exemplary usage of this function is given in the code snipped below, along with the default output.

\begin{Schunk}
\begin{Sinput}
R> res <- svsample(ret, priormu = c(-10, 1), priorphi = c(20, 1.1),
+    priorsigma = 0.1)
\end{Sinput}
\end{Schunk}

\begin{Soutput}
Calling GIS_C MCMC sampler with 11000 iter. Series length is 3139.

  0% [+++++++++++++++++++++++++++++++++++++++++++++++++++] 100%

Timing (elapsed): 12.92 seconds.
851 iterations per second.

Converting results to coda objects... Done!
Summarizing posterior draws... Done!
\end{Soutput}

As can be seen, this function calls the main MCMC sampler and converts its output to \pkg{coda}-compatible objects. The latter is done mainly for reasons of compatibility and in order to have straightforward access to the convergence diagnostics checks implemented there. Moreover, some summary statistics for the posterior draws are calculated. The return value of \code{svsample} is an object of type \code{svdraws} which is a named list with eight elements, holding (1) the parameter draws in \code{para}, (2) the latent log-volatilities in \code{latent}, (3) the initial latent log-volatility draw in \code{latent0}, (4) the data provided in \code{y}, (5) the sampling runtime in \code{runtime}, (6) the prior hyperparameters in \code{priors}, (7) the thinning values in \code{thinning}, and (8) summary statistics of these draws, alongside some common transformations thereof, in \code{summary}.

 \subsection{Assessing the output and displaying the results}
 \label{check}

 Following common practice, \code{print} and \code{summary} methods are available for \code{svdraws} objects. Each of these has two optional parameters, \code{showpara} and \code{showlatent}, specifying which output should be displayed. If \code{showpara} is \code{TRUE} (the default), values/summaries of the parameter draws are shown. If \code{showlatent} is \code{TRUE} (the default), values/summaries of the latent variable draws are shown. In the example below, the summary for the parameter draws only is displayed.

\begin{Schunk}
\begin{Sinput}
R> summary(res, showlatent = FALSE)
\end{Sinput}
\begin{Soutput}
Summary of 10000 MCMC draws after a burn-in of 1000.
Prior distributions:
mu        ~ Normal(mean = -10, sd = 1)
(phi+1)/2 ~ Beta(a0 = 20, b0 = 1.1)
sigma^2   ~ 0.1 * Chisq(df = 1)

Posterior draws of parameters (thinning = 1):
              mean      sd       5%      50%     95%  ESS
mu        -10.1366 0.22711 -10.4749 -10.1399 -9.7933 4552
phi         0.9935 0.00282   0.9886   0.9938  0.9977  397
sigma       0.0656 0.01001   0.0509   0.0649  0.0830  143
exp(mu/2)   0.0063 0.00075   0.0053   0.0063  0.0075 4552
sigma^2     0.0044 0.00139   0.0026   0.0042  0.0069  143
\end{Soutput}
\end{Schunk}

There are several plotting functions specifically designed for objects of class \code{svsample} which are described in the following paragraphs.
 \begin{enumerate}
  \item[(1)] \code{volplot}: Plots posterior quantiles of the latent volatilities in percent, i.e., empirical quantiles of the posterior distribution of $100\exp(h_t/2)$, over time. Apart from the mandatory \code{svsample}-object itself, this function takes several optional arguments. Only some are mentioned here; for an exhaustive list please see the corresponding help document accessible through \code{?volplot} or \code{help(volplot)}. Selected optional arguments that are commonly used include \code{forecast} for $n$-step-ahead volatility prediction, \code{dates} for labels on the $x$-axis, alongside some graphical parameters. The code snipped below shows a typical example and Figure~\ref{fig3} displays its output.

\begin{Schunk}
\begin{Sinput}
R> volplot(res, forecast = 100, dates = exrates$date[-1])
\end{Sinput}
\end{Schunk}

\begin{figure}[!htp]
\begin{center}
 \includegraphics[width=\textwidth]{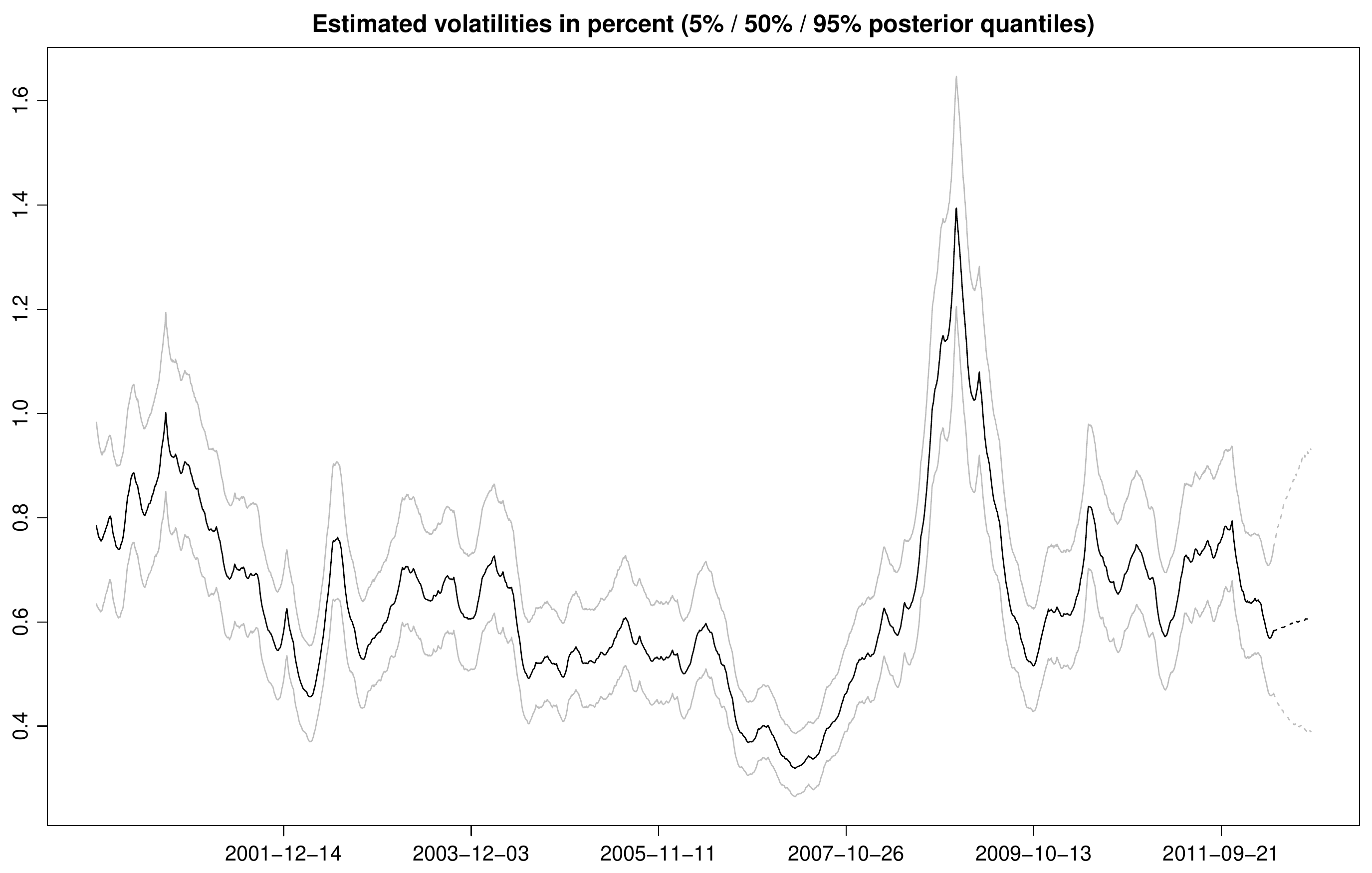}
 \caption{Visualization of estimated contemporaneous volatilities of EUR-USD exchange rates, as provided by \code{volplot}. If not specified otherwise, posterior medians and 5\%/95\% quantiles are plotted. The dotted lines on the right indicate predicted future volatilities.}
 \label{fig3}
\end{center}
\end{figure}

In case the user wants to display different posterior quantiles, the \code{updatesummary} function has to be called first. See the code below for an example and Figure~\ref{fig4} for the corresponding plot.
 
\begin{Schunk}
\begin{Sinput}
R> res <- updatesummary(res, quantiles = c(0.01, 0.1, 0.5, 0.9, 0.99))
R> volplot(res, forecast = 100, dates = exrates$date[-1])
\end{Sinput}
\end{Schunk}

\begin{figure}[t]
\begin{center}
 \includegraphics[width=\textwidth]{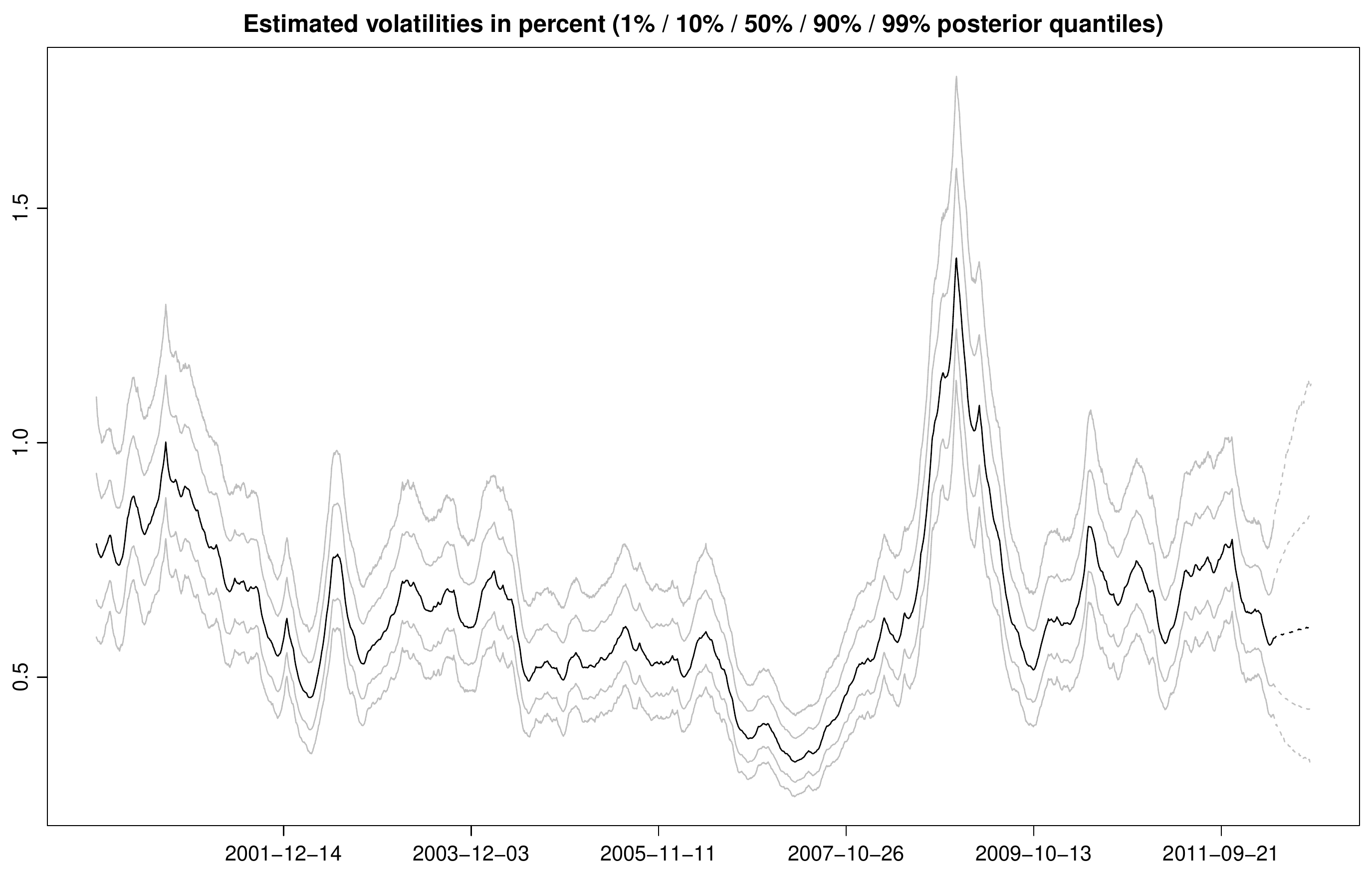}
 \caption{As above, now with medians (black line) and 1\%/10\%/90\%/99\% quantiles (gray lines). This behavior can be achieved through a preceding call of \code{updatesummary}.}
 \label{fig4}
\end{center}
\end{figure}

 \item[(2)] \code{paratraceplot}: Displays trace plots for the parameters contained in $\allpara$. Note that the burn-in has already been discarded. Figure~\ref{fig5} shows an example.

\begin{Schunk}
\begin{Sinput}
R> par(mfrow = c(3, 1))
R> paratraceplot(res)
\end{Sinput}
\end{Schunk}
 
\begin{figure}[!htp]
\begin{center}
 \includegraphics[width=\textwidth]{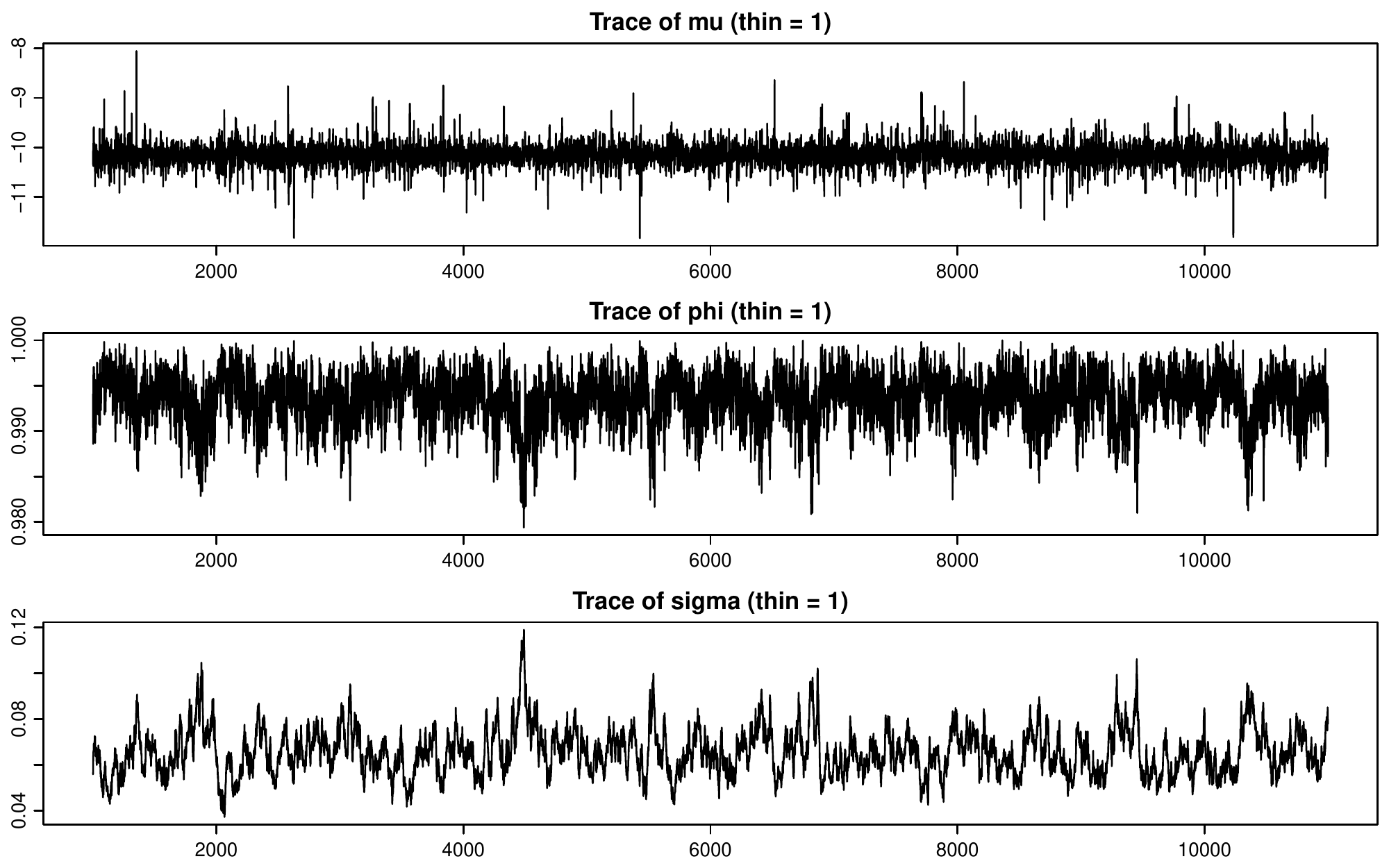}
 \caption{Trace plots of posterior draws for the parameters $\mupar, \phipar, \sigmapar$.}
 \label{fig5}
\end{center}
 \end{figure}

\item[(3)] \code{paradensplot}: Displays a kernel density estimate for the parameters contained in $\allpara$. If the argument \code{showobs} is \code{TRUE} (which is the default), individual posterior draws are indicated through a rug, i.e., short vertical lines along the $x$-axis. For quicker drawing of large posterior samples, this argument should be set to \code{FALSE}. If the argument \code{showprior} is \code{TRUE} (which is the default), the prior distribution is indicated through a dashed gray line. Figure~\ref{fig6} shows a sample output for the EUR-USD exchange rates obtained from the \code{exrates} dataset.

\begin{Schunk}
\begin{Sinput}
R> par(mfrow = c(1, 3))
R> paradensplot(res, showobs = FALSE)
\end{Sinput}
\end{Schunk}

\begin{figure}[!htp]
\begin{center}
 \includegraphics[width=\textwidth]{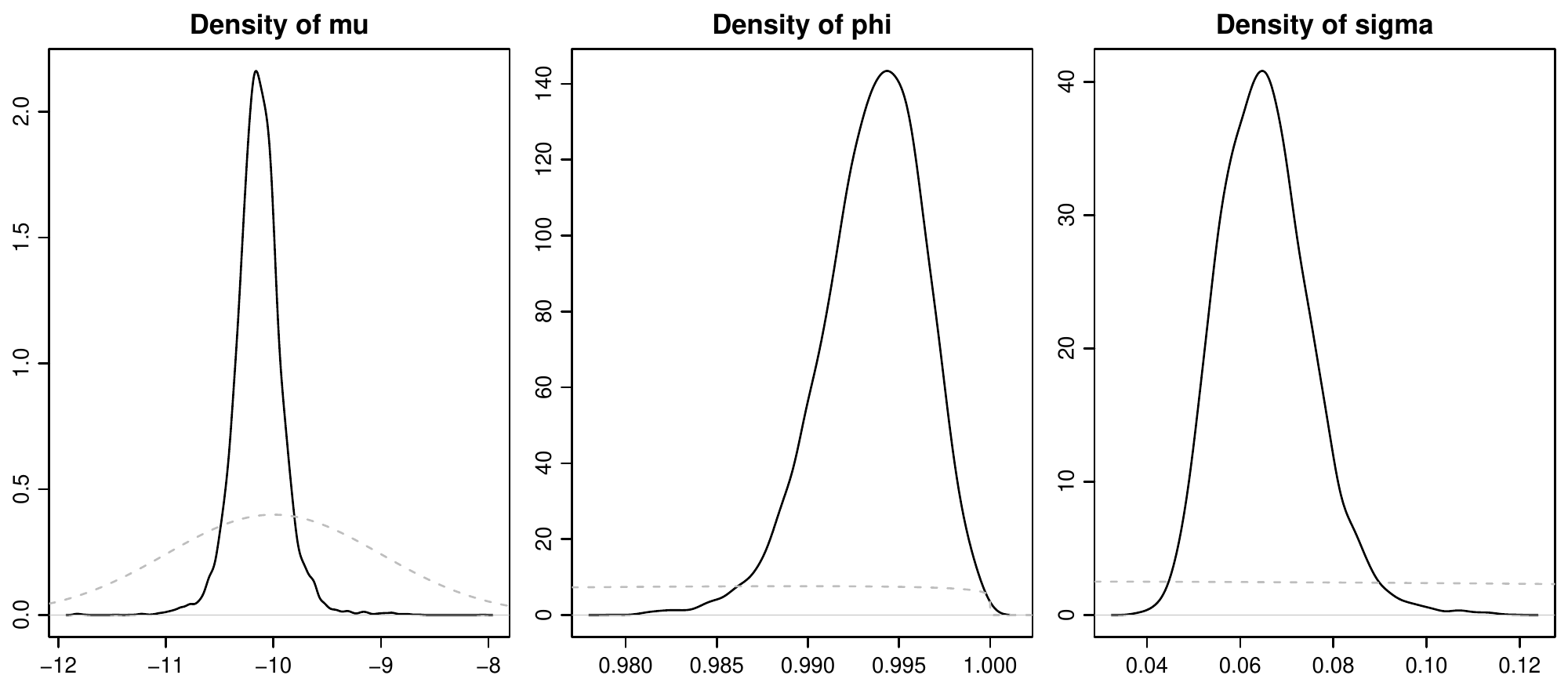}
 \caption{Posterior density estimates (black solid lines) along with prior densities (dashed gray lines). Individual posterior draws are indicated by the underlying rug.}
 \label{fig6}
\end{center}
\end{figure}
\end{enumerate}

\begin{figure}[!ht]
\begin{center}
 \includegraphics[width=\textwidth]{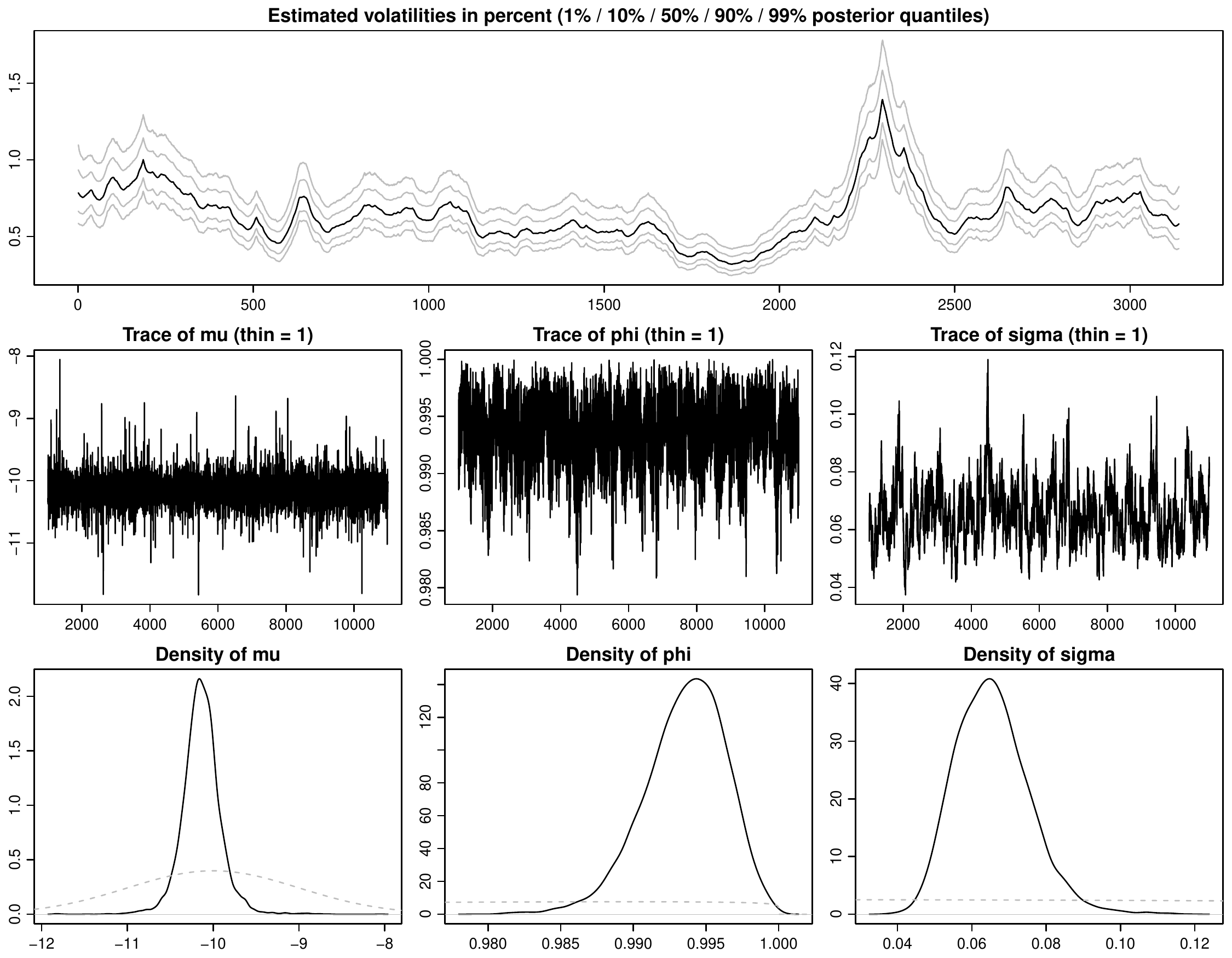}
 \caption{Illustration of the default \code{plot} method for \code{svdraws}-objects. This visualization combines \code{volplot} (Figure~\ref{fig4}), \code{traceplot} (Figure~\ref{fig5}), and \code{paradensplot} (Figure~\ref{fig6}) into one single plot.}
 \label{fig7}
\end{center}
\end{figure}

The generic \code{plot} method for \code{svdraws} objects combines all above plots into one plot. All arguments described above can be used. See \code{?plot.svsample} for an exhaustive summary of possible arguments and Figure~\ref{fig7} for an example.

\begin{Schunk}
\begin{Sinput}
R> plot(res, showobs = FALSE)
\end{Sinput}
\end{Schunk}

\begin{figure}[!ht]
\begin{center}
 \includegraphics[width=\textwidth]{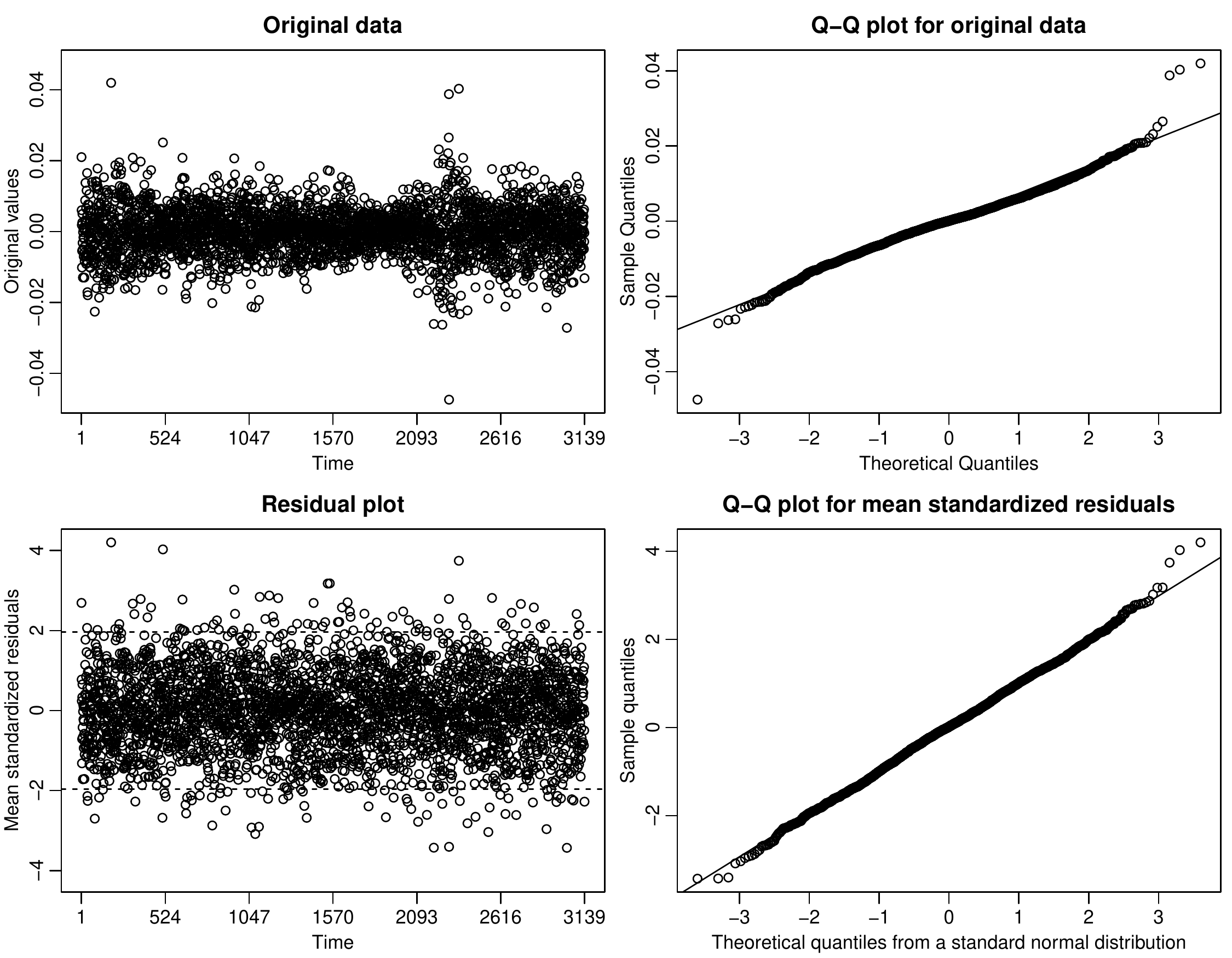}
 \caption{Mean standardized residual plots for assessing the model fit, as provided by the corresponding \code{plot} method. The dashed lines in the bottom left panel indicate the 2.5\%/97.5\% quantiles of the standard normal distribution.}
 \label{fig8}
\end{center}
\end{figure}

For extracting standardized residuals, the \code{residuals}/\code{resid} method can be used on a given \code{svdraws} object. With the optional argument \code{type}, the type of summary statistic may be specified. Currently, \code{type} is allowed to be either \code{"mean"} or \code{"median"}, where the former corresponds to the default value. This method returns a real vector of class \code{svresid} which contains the requested summary statistic of standardized residuals for each point in time. There is also a \code{plot} method available, providing the option of comparing the standardized residuals to the original data when given through the argument \code{origdata}. See the code below for an example and Figure~\ref{fig8} for the corresponding output.

\begin{Schunk}
\begin{Sinput}
R> myresid <- resid(res)
R> plot(myresid, ret)
\end{Sinput}
\end{Schunk}

\section[Using stochvol within other samplers]{Using \pkg{stochvol} within other samplers}
\label{other}
We demonstrate how the \pkg{stochvol} package can be used to incorporate stochastic volatility into any given MCMC sampler. This is particularly easy when the sampler itself is coded in \proglang{R} or \proglang{C}/\proglang{C++} and applies, e.g., to many of the specialized procedures listed in the CRAN task view about Bayesian inference, available at \url{http://cran.r-project.org/web/views/Bayesian.html}. For the sake of simplicity, we explain the procedure using a ``hand-coded'' Gibbs-sampler for the Bayesian normal linear model with $n$ observations and $k=p-1$ predictors, given through
\begin{eqnarray}
 \label{reg}
\yb|\betab,\Sigmab \sim \Normal{\Xb\betab, \Sigmab}.
\end{eqnarray}
Here, $\yb$ denotes the $n\times1$ vector of responses, $\Xb$ is the $n \times p$ design matrix containing ones in the first column and the predictors in the others, and $\betab = (\beta_0, \beta_1, \dots, \beta_{p-1})^\top$ stands for the $p\times 1$ vector of regression coefficients. In the following sections, we discuss two specifications of the $n\times n$ error covariance matrix $\Sigmab$.

\subsection{The Bayesian normal linear model with homoskedastic errors}
\label{homo}
The arguably simplest specification of the error covariance matrix in Equation~\ref{reg} is given by $\Sigmab\equiv\sigmaepsextrasuper2\Ib$, where $\Ib$ denotes the $n$-dimensional unit matrix. This specification is used in many applications and commonly referred to as the linear regression model with homoskedastic errors. To keep things simple, let model parameters $\betab$ and $\sigmaepsextrasuper2$ be equipped with the usual conjugate prior $p(\betab, \sigmaepsextrasuper2) = p(\betab|\sigmaepsextrasuper2)p(\sigmaepsextrasuper2)$, where
\begin{eqnarray*}
\betab|\sigmaepsextrasuper2 &\sim& \Normal{\bb0, \sigmaepsextrasuper2\Bb0},\\
 \sigmaepsextrasuper2&\sim& \Gammainv{c_0, C_0}.
\end{eqnarray*}

Commonly, samples from the posterior distribution of this model are obtained through a Gibbs-algorithm, where draws are generated in turn from the full conditional distributions $\betab|\yb,\sigmaepsextrasuper2 \sim \Normal{\bpostb, \Bpostb}$ with
\[
\bpostb = \left(\XtX + \Bb0^{-1} \right)^{-1}\left( \Xb^\top\yb + \Bb0^{-1}\bb0 \right), \qquad
\Bpostb = \sigmaepsextrasuper2\left(\XtX + \Bb0^{-1} \right)^{-1},
\]

and $\sigmaepsextrasuper2|\yb,\betab \sim \Gammainv{c_n, C_n}$ with
\[
c_n = c_0 + \frac{n}{2} + \frac{p}{2}, \qquad
C_n = C_0 + \frac{1}{2}
 \left(
 (\yb - \Xb\betab)^\top(\yb - \Xb\betab) +
 (\betab - \bb0)^\top\Bb0^{-1}(\betab - \bb0)
 \right).
\]

In \proglang{R}, this can straightforwardly be coded as follows:
\begin{itemize}
 \item Set the seed to make results reproducible and simulate some data from the underlying model:
\begin{Schunk}
\begin{Sinput}
R> set.seed(123456)
R> n <- 1000
R> beta.true <- c(0.1, 0.5)
R> sigma.true <- 0.01
R> X <- matrix(c(rep(1, n), rnorm(n, sd = sigma.true)), nrow = n)
R> y <- rnorm(n, X %*% beta.true, sigma.true)
\end{Sinput}
\end{Schunk}
 
\item Specify the size of the burn-in and the number of draws thereafter; set the prior parameters:
\begin{Schunk}
\begin{Sinput}
R> burnin <- 100
R> draws <- 5000
R> b0 <- matrix(c(0, 0), nrow = ncol(X))
R> B0inv <- diag(c(10^-10, 10^-10))
R> c0 <- 0.001
R> C0 <- 0.001
\end{Sinput}
\end{Schunk}
 \item Pre-calculate some values outside the main MCMC loop:
\begin{Schunk}
\begin{Sinput}
R> p <- ncol(X)
R> preCov <- solve(crossprod(X) + B0inv)
R> preMean <- preCov %*% (crossprod(X, y) + B0inv %*% b0)
R> preDf <- c0 + n/2 + p/2
\end{Sinput}
\end{Schunk}
\item Assign some storage space for holding the draws and set an initial value for $\sigmaepsextrasuper2$:
\begin{Schunk}
\begin{Sinput}
R> draws1 <- matrix(NA_real_, nrow = draws, ncol = p + 1)
R> colnames(draws1) <- c(paste("beta", 0:(p-1), sep = "_"), "sigma")
R> sigma2draw <- 1
\end{Sinput}
\end{Schunk}
\item Run the main sampler: Iteratively draw from the conditional bivariate Gaussian distribution $\betab|\yb,\sigmaepsextrasuper2$, e.g., through the use of \pkg{mvtnorm} \citep{r:mvt}, and the conditional Inverse Gamma distribution $\sigmaepsextrasuper2|\yb,\betab$.
\begin{Schunk}
\begin{Sinput}
R> for (i in -(burnin-1):draws) {
+    betadraw <- as.numeric(mvtnorm::rmvnorm(1, preMean,
+      sigma2draw * preCov))
+    tmp <- C0 + 0.5 * (crossprod(y - X %*% betadraw) +
+      crossprod((betadraw - b0), B0inv) %*% (betadraw - b0))
+    sigma2draw <- 1 / rgamma(1, preDf, rate = tmp)
+    if (i > 0) draws1[i, ] <- c(betadraw, sqrt(sigma2draw))
+  }
\end{Sinput}
\end{Schunk}
\item Calculate posterior means in order to obtain point estimates for the parameters:
\begin{Schunk}
\begin{Sinput}
R> colMeans(draws1)
\end{Sinput}
\begin{Soutput}
    beta_0     beta_1      sigma 
0.09991649 0.50472433 0.01027775 
\end{Soutput}
\end{Schunk}

\item Visualize the draws through \pkg{coda}'s native \code{plot} method:

\begin{Schunk}
\begin{Sinput}
R> plot(coda::mcmc(draws1), show.obs = FALSE)
\end{Sinput}
\end{Schunk}

\begin{figure}[!ht]
\begin{center}
\includegraphics[width=\textwidth]{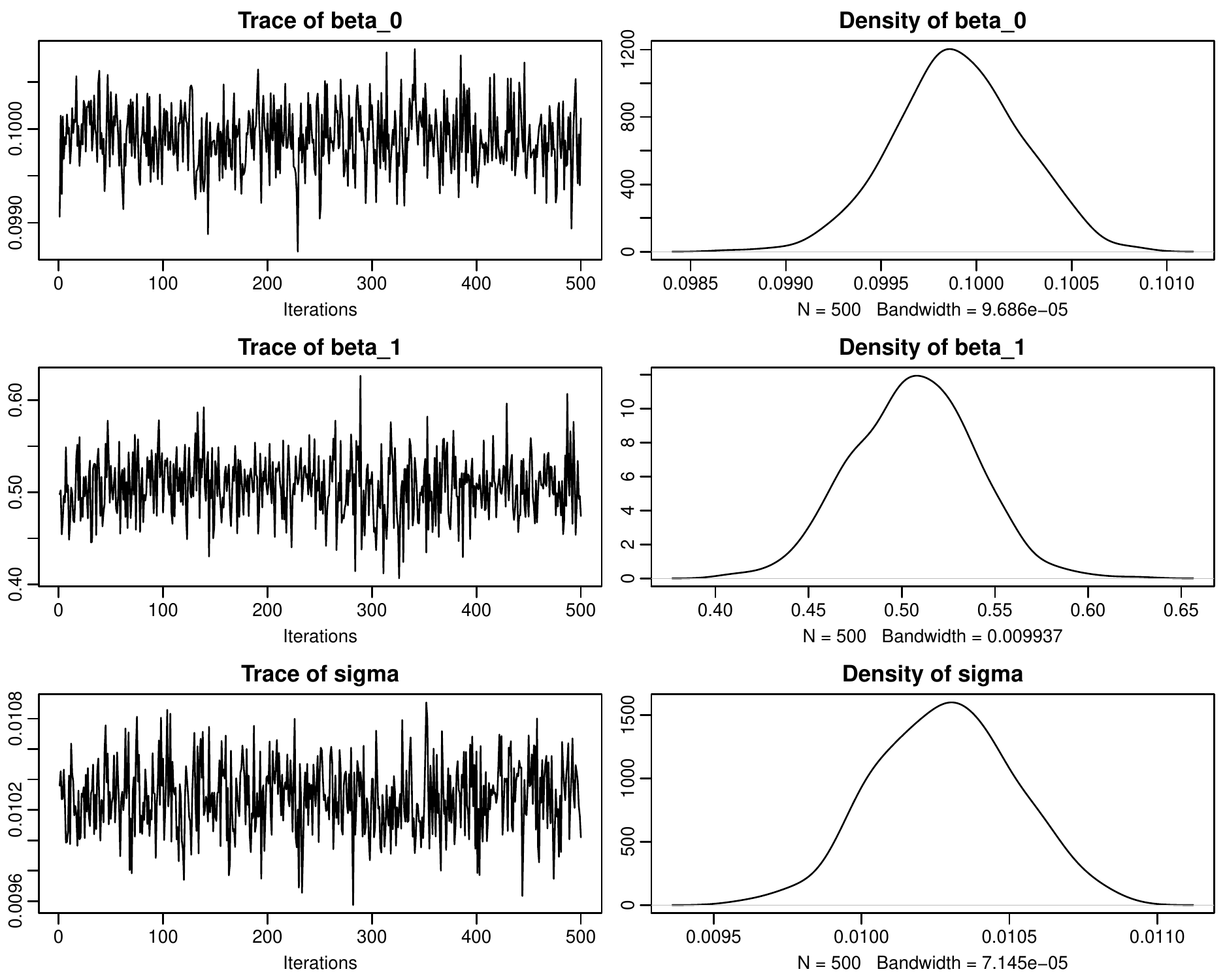}
\caption{Trace plots and kernel density estimates for some draws from the marginal posterior distributions in the regression model with heteroskedastic errors. Underlying data is simulated with $\betab^\text{true}=(0.1, 0.5)^\top$, $\sigmaepsextrasuper{\text{true}} = 0.01$, $n=1000$.}
\label{homodraws}
\end{center}
\end{figure}

\end{itemize}

\subsection{The Bayesian normal linear model with SV errors}
\label{hetero}

Instead of homoskedastic errors, we now specify the error covariance matrix in Equation~\ref{reg} to be $\Sigmab \equiv \diag{\e^{h_1}, \dots, \e^{h_n}}$, thus introducing nonlinear dependence between the observations due to the AR(1)-nature of $\hb$. Instead of cooking up an entire new sampler, we adapt the code from above utilizing the \pkg{stochvol} package. To do so, we simply replace the sampling step of $\sigmaepsextrasuper{2}$ from an Inverse-Gamma distribution with a sampling step of $\allpara$ and $\hb$ through a call to \code{svsample2}.\footnote{In earlier version of the \pkg{stochvol} package, this function was called \code{.svsample}.} This function is a minimal-overhead version of the regular \code{svsample}. It provides the full sampling functionality of the original version but has slightly different default values, a simplified return value structure, and it does not perform costly input checks. Thus, it executes faster and is more suited for repeated calls. The drawback is that it needs to be used with proper care.\footnote{Erroneous or incompatible input values most likely result in run-time errors of compiled code, often implying a segmentation fault and the consecutive abnormal termination of \proglang{R}. This can render debugging tedious.} Note that the current draws of the variables need to be passed to the function through \code{startpara} and \code{startlatent}.

Here is how it goes:
\begin{itemize}
 \item Simulate some data:
\begin{Schunk}
\begin{Sinput}
R> mu.true <- log(sigma.true^2)
R> phi.true <- 0.97
R> vv.true <- 0.3
R> simresid <- svsim(n, mu = mu.true, phi = phi.true, sigma = vv.true)
R> y <- X %*% beta.true + simresid$y
\end{Sinput}
\end{Schunk}
 \item Specify configuration parameters and prior values:
\begin{Schunk}
\begin{Sinput}
R> draws <- 50000
R> burnin <- 1000
R> thinning <- 10
R> priormu <- c(-10, 2)
R> priorphi <- c(20, 1.5)
R> priorsigma <- 1
\end{Sinput}
\end{Schunk}
 \item Assign some storage space for holding the draws and set initial values:
\begin{Schunk}
\begin{Sinput}
R> draws2 <- matrix(NA_real_, nrow = floor(draws / thinning),
+    ncol = 3 + n + p)
R> colnames(draws2) <- c("mu", "phi", "sigma",
+    paste("beta", 0:(p-1), sep = "_"), paste("h", 1:n, sep = "_"))
R> betadraw <- c(0, 0)
R> svdraw <- list(para = c(mu = -10, phi = 0.9, sigma = 0.2),
+    latent = rep(-10, n))
\end{Sinput}
\end{Schunk}
 \item Run the main sampler, i.e., iteratively draw
  \begin{itemize}
   \item the latent volatilities/parameters by conditioning on the regression parameters and calling \code{svsample2},
   \item the regression parameters by conditioning on the latent volatilities and calling \code{rmvnorm}:
  \end{itemize}
\begin{Schunk}
\begin{Sinput}
R> for (i in -(burnin-1):draws) {
+    ytilde <- y - X %*% betadraw
+    svdraw <- svsample2(ytilde, startpara = para(svdraw),
+      startlatent = latent(svdraw), priormu = priormu,
+      priorphi = priorphi, priorsigma = priorsigma)
+    normalizer <- as.numeric(exp(-latent(svdraw) / 2))
+    Xnew <- X * normalizer
+    ynew <- y * normalizer
+    Sigma <- solve(crossprod(Xnew) + B0inv)
+    mu <- Sigma %*% (crossprod(Xnew, ynew) + B0inv %*% b0)
+    betadraw <- as.numeric(mvtnorm::rmvnorm(1, mu, Sigma))
+    if (i > 0 & i %% thinning == 0) { 
+      draws2[i/thinning, 1:3] <- para(svdraw)
+      draws2[i/thinning, 4:5] <- betadraw
+      draws2[i/thinning, 6:(n+5)] <- latent(svdraw)
+    }
+  }
\end{Sinput}
\end{Schunk}

\item Finally, visualize and summarize (some) posterior draws:

\begin{Schunk}
\begin{Sinput}
R> plot(coda::mcmc(draws2[, 4:7]), show.obs = FALSE)
\end{Sinput}
\end{Schunk}

\begin{figure}[!ht]
 \begin{center}
\includegraphics[width=\textwidth]{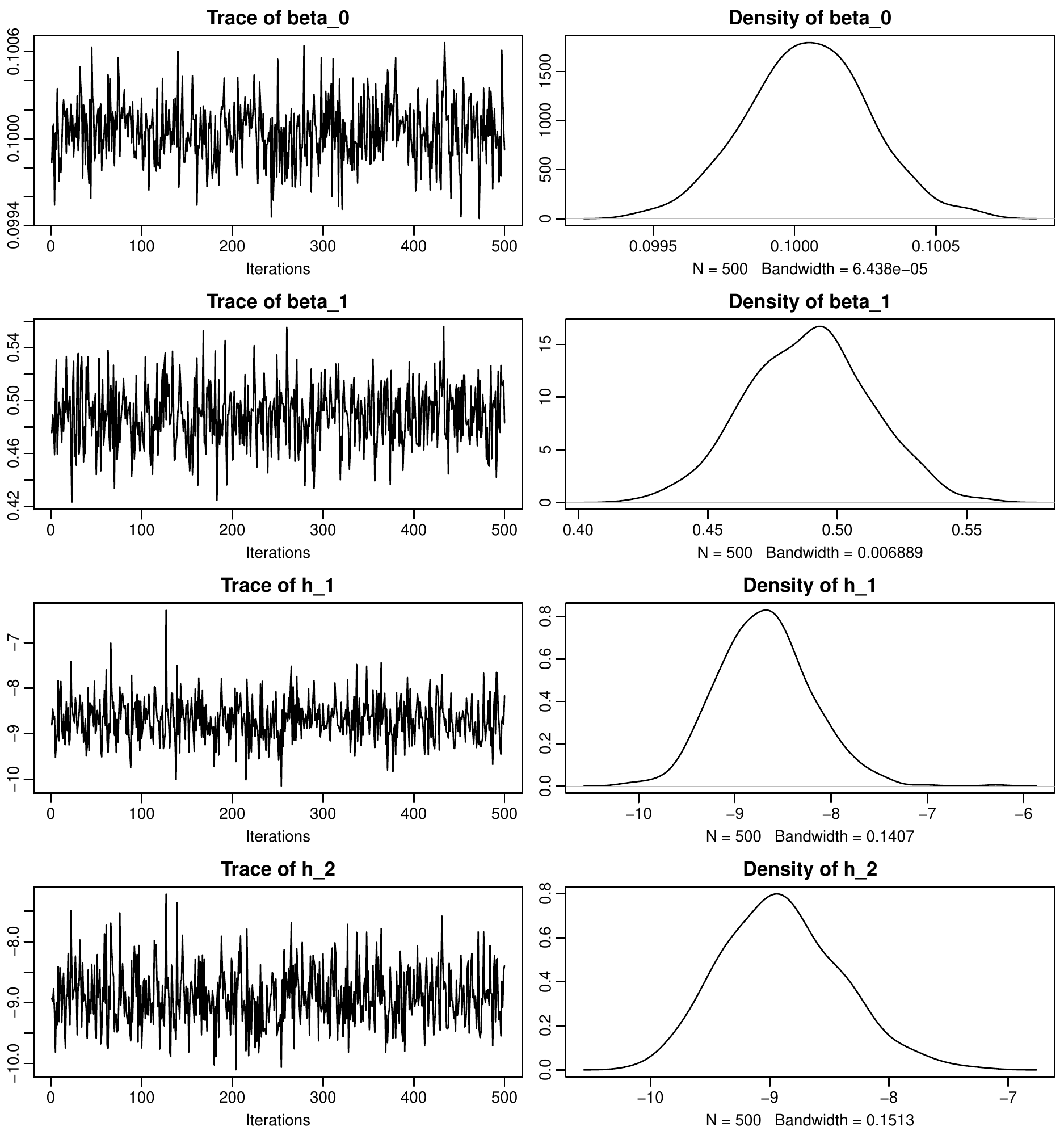}
\caption{Trace plots and kernel density estimates
in the regression model with heteroskedastic errors. Data is simulated with $\betab^\text{true}=(0.1, 0.5)^\top$, $h_1^\text{true} = -8.28$, $h_2^\text{true} = -8.50$, $n=1000$.}
\label{heterodraws}
 \end{center}
\end{figure}

\begin{Schunk}
\begin{Sinput}
R> colMeans(draws2[, 4:8])
\end{Sinput}
\end{Schunk}

\begin{Schunk}
\begin{Soutput}
    beta_0     beta_1        h_1        h_2        h_3 
 0.1000519  0.4873449 -8.6739795 -8.8931827 -9.0834209 
\end{Soutput}
\end{Schunk}

\end{itemize}

It should be noted that even though \code{svsample2} is considerably faster than \code{svsample}, the cost of interpreting this function in each MCMC iteration is still rather high (which lies in the nature of \proglang{R} as an interpreted language). Thus, as of package version \code{0.8-0}, a single-step \code{update} is also made available at the \proglang{C}/\proglang{C++} level for samplers coded there. For details, please consult the \code{NEWS} file in the package source; for an application case using this approach, see \cite{kas-etal:ana}.

\section{Illustrative predictive exercise}
In the following, we compare the performance of the Bayesian normal linear model with homoskedastic errors from Section~\ref{homo} with the Bayesian normal linear model with SV errors from Section~\ref{hetero} using the \code{exrates} dataset introduced in Section~\ref{prep}. As a benchmark, we also include the Bayesian normal linear model with GARCH(1,1) errors given through Equation~\ref{reg} with 
\begin{eqnarray*}
 \label{garch1} \Sigmab &=& \diag{\sigma^2_1,\dots,\sigma^2_n},\\
 \label{garch2} \sigma^2_t &=& \alpha_0+\alpha_1\tilde y_{t-1}^2+\alpha_2\sigma_{t-1}^2,
\end{eqnarray*}
where time index $t=1,\dots,n$. In the second equation, $\tilde y_{t-1}$ denotes the past ``residual'', i.e., the $(t-1)$th element of $\tilde \yb = \yb - \Xb\betab$.

\subsection{Model setup}

We again use the daily price of 1 EUR in USD from January 3, 2000 to April 4, 2012, denoted by $\pb = (p_1, p_2, \dots, p_n)^\top$. This time however, instead of modeling log returns, we investigate the development of log levels by regression. Let $\yb$ contain the logarithm of all observations except the very first one, and let $\Xb$ denote the design matrix containing ones in the first column and lagged log prices in the second, i.e.,
\begin{eqnarray*}
\yb = 
\begin{pmatrix}
 \log p_2\\
 \log p_3\\
 \vdots\\
 \log p_n
\end{pmatrix}, \quad
\Xb =
\begin{pmatrix}
1&\log p_1\\
1&\log p_2\\
\vdots&\vdots\\
1&\log p_{n-1}
\end{pmatrix}.
\label{regression}
\end{eqnarray*}
Note that this specification simply corresponds to an AR(1) model for the log prices with three different error specifications: homoskedastic, SV, GARCH(1,1). It is a generalization of directly dealing with demeaned log returns where $\beta_0$ is a priori assumed to be equal to the mean of the log returns and $\beta_1$ is held constant at $1$.

\begin{figure}[!ht]
\begin{center}
 \includegraphics[width=\textwidth]{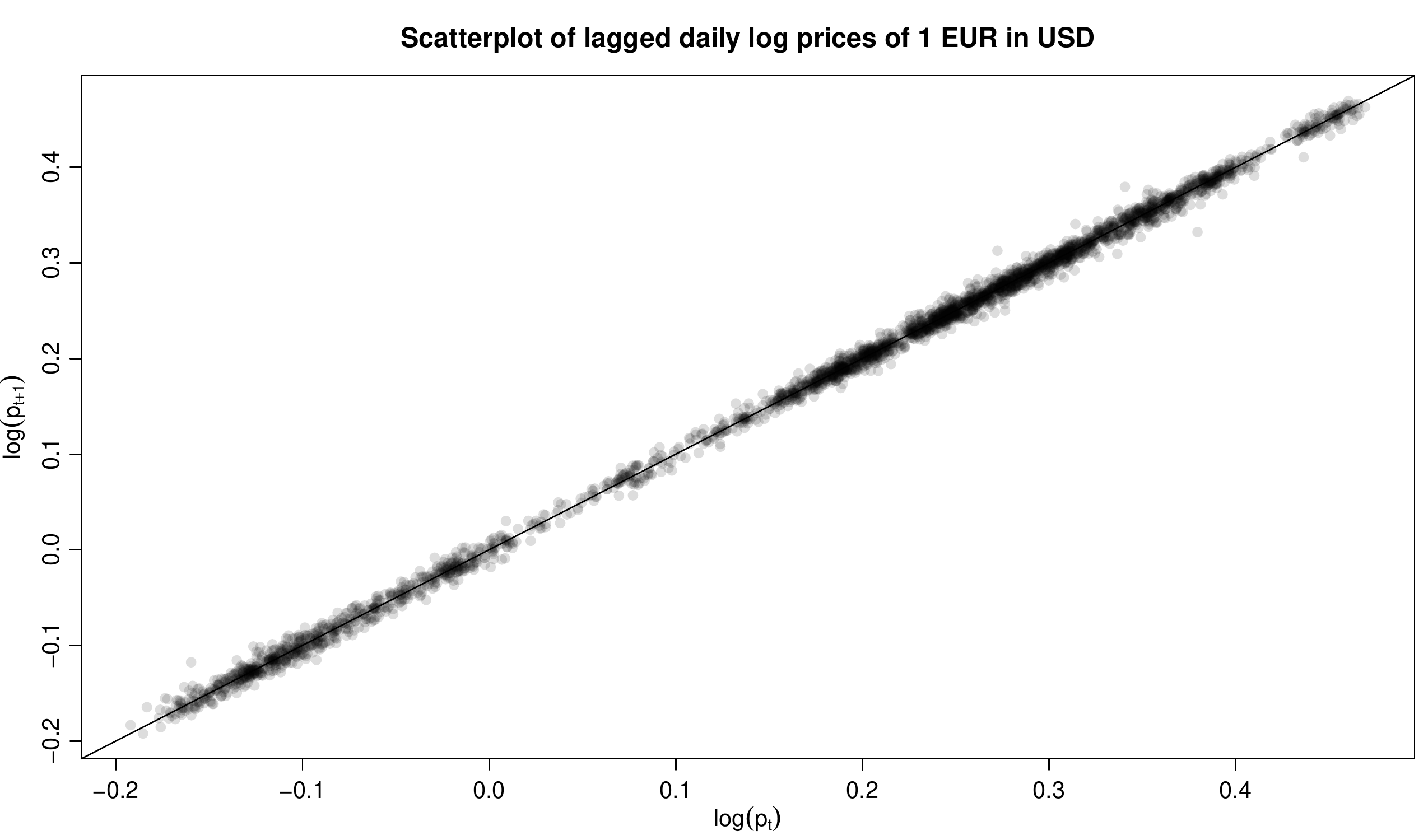}
\end{center}
\caption{Scatterplot of daily log prices at time $t$ against daily log prices at time $t+1$. The solid line indicates the identity function $f(x) = x$.}
\label{datascatter}
\end{figure}

Irrespectively of the error specification, we expect the posterior distribution of $\betab$ to spread around $(0,1)^\top$ which corresponds to a random walk. A scatterplot of $\log p_t$ against $\log p_{t+1}$, displayed in Figure~\ref{datascatter}, confirms this.

\subsection{Posterior inference}

To obtain draws from the posterior distributions for the models with homoskedastic/SV errors, samplers developed in Section~\ref{other} are used. For the GARCH-errors, a simple random walk Metropolis-Hastings sampler is employed. We run each of the three samplers for 100\,000 iterations and discard the first 10\,000 draws as burn-in. Starting values and tuning parameters are set using maximum likelihood estimates obtained through the \proglang{R} package \pkg{fGarch} \citep{r:fGarch}.

Aiming for comparable results with minimal prior impact on predictive performance, the hyperparameters are chosen to yield vague priors: $\bb{0} = (0, 0)^\top$, $\Bb{0} = \diag{10^{10}, 10^{10}}$, $c_0 = C_0 = 0.001$, $b_\mu = 0$, $B_\mu=10^4$, $a_0=1$, $b_0 = 1$, $B_\sigmapar=1$. For the GARCH(1,1) parameter vector $\bm{\alpha} = (\alpha_0, \alpha_1, \alpha_2)^\top$, we pick independent flat priors on $\mathbb{R}^+$ for the components; the initial variance $\sigma_0^2$ is fixed to the empirical residual variance and $\tilde \y_0$ is assumed to be zero. Due to the length of the dataset (and its obvious heteroskedasticity), none of these specific choices seem to particularly influence the following analysis. Note, however, that for shorter series or series with less pronounced heteroskedasticity, sensible prior choices are of great importance as the likelihood carries less information in these cases.

\begin{figure}[!ht]
\begin{center}
 \includegraphics[width=\textwidth]{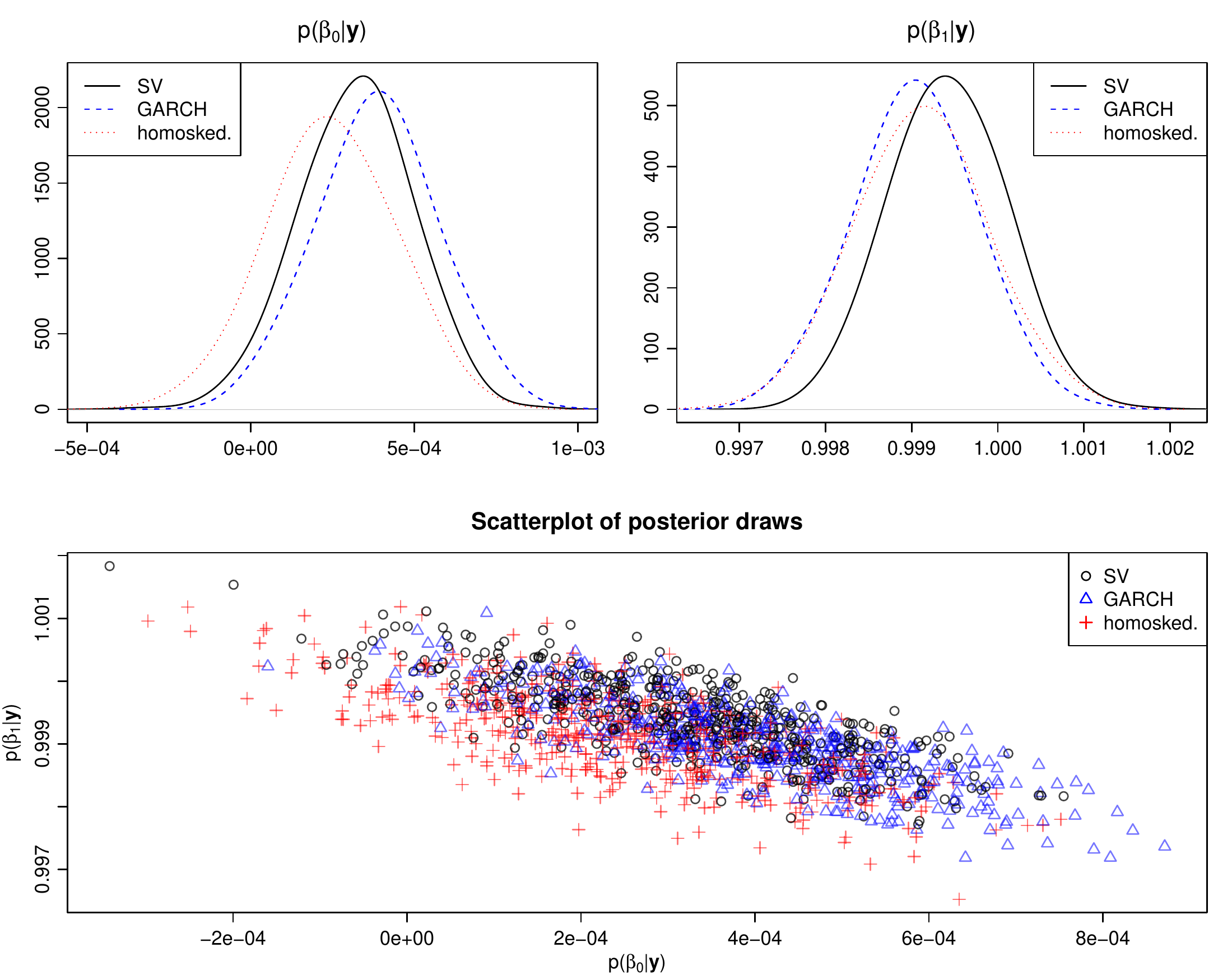}
 \caption{Visualization of the posterior distributions $\betab|\yb$ for the model with SV regression residuals, the model with GARCH(1,1) regression residuals, and the model with homoskedastic regression residuals. Top panels: Kernel density estimates of the univariate posterior marginal distributions. Bottom panel: Bivariate scatterplot of posterior draws.}
  \label{betapost}
\end{center}
 \end{figure}

The samplers yield slightly different posteriors for $\betab$, visualized in Figure~\ref{betapost}. In the top panels, the estimated marginal posterior densities $p(\beta_0|\yb)$ and $p(\beta_1|\yb)$ are displayed; the bottom panel depicts a scatterplot of draws from the joint posterior $\betab|\yb$.

\begin{figure}[!ht]
 \begin{center}
  \includegraphics[width=\textwidth]{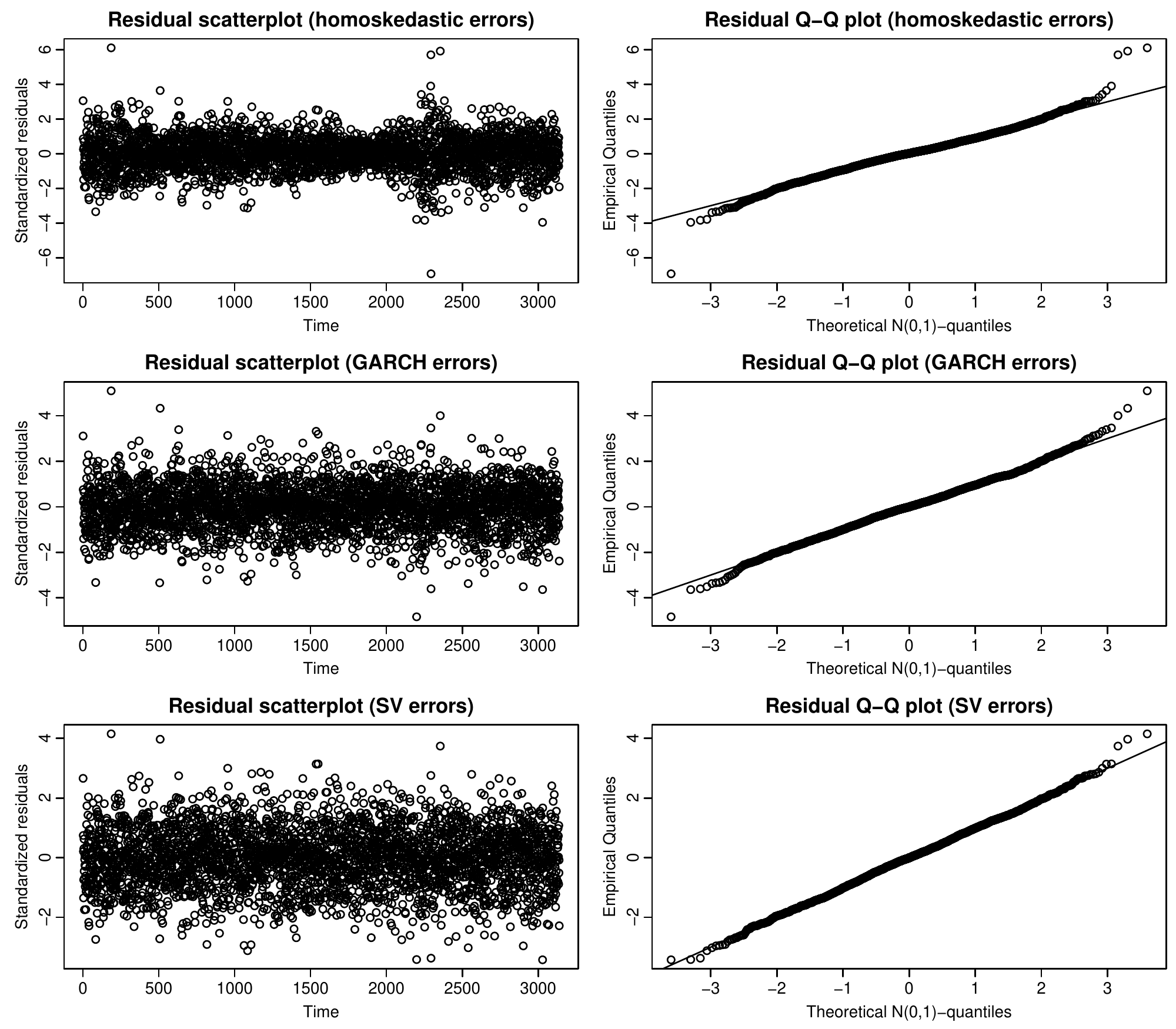}
 \end{center}
 \caption{Visualization of mean standardized residuals. Top left panel shows a scatterplot against time for the model with homoskedastic errors, middle left panel shows this plot for the model with GARCH errors, and bottom left panel shows this plot for the model with SV errors. Quantile-Quantile plots of empirical quantiles against expected quantiles from a $\Normal{0,1}$-distribution are displayed on the panels on the right-hand side.}
 \label{qqplot}
\end{figure}

To assess the model fit, mean standardized residuals are depicted in Figure~\ref{qqplot}. The model with homoskedastic errors shows deficiencies in terms of pronounced dependence amongst the residuals. This can clearly be seen in the top left panel, where mean standardized residuals are plotted against time. The middle left panel shows the same plot for the model with GARCH errors where this effect is greatly diminished. The bottom left panel pictures that plot for the model with SV errors; here, this effect practically vanishes. Moreover, in the model with homoskedastic errors, the normality assumption about the unconditional error distribution is clearly violated. This can be seen by inspecting the quantile-quantile plot in the top right panel, where observed residuals exhibit much heavier tails than one would expect from a normal distribution. The model with GARCH errors provides a better fit, however, heavy tails are still visible. Standardized residuals from the model with SV errors appear to be approximately normal with only few potential outliers.

\subsection{Measuring predictive performance and model fit}
Within a Bayesian framework, a natural way of assessing the predictive performance of a given model is through its \emph{predictive density} (sometimes also referred to as \emph{posterior predictive distribution}). It is given through
\begin{eqnarray}
 \label{preddens}
p(\y_{t+1}|\yb^o_{[1:t]}) = \int\limits_{\bm K} \!
 p(y_{t+1}|\yb^o_{[1:t]}, \unobs) \times
 p(\unobs|\yb^o_{[1:t]}) 
 \, \dif \unobs,
\end{eqnarray}
where $\unobs$ denotes the vector of all unobservables, i.e., parameters and possible latent variables.
Note that for the model with homoskedastic errors, $\unobs=(\betab,\sigmaeps)^\top$; for the model with SV errors, $\unobs=(\betab, \allpara, \hb)^\top$; in the GARCH(1,1) case, $\unobs=(\betab,\bm{\alpha},\sigma_0^2,\tilde y_0^2)^\top$. By using the superscript $o$ in $\yb^o_{[1:t]}$, we follow \cite{gew-ami:com} and denote \emph{ex post} realizations (observations) for the set of points in time $\{1,2,\dots ,t\}$ of the \emph{ex ante} random values $\yb_{[1:t]} = (\y_1, \y_2, \dots,\y_t)^\top$. Integration is carried out over ${\bm K}$ which simply stands for the space of all possible values for $\unobs$. Equation~\ref{preddens} can be viewed as the integral of the likelihood function over the joint posterior distribution of the unobservables $\unobs$. Thus, it can be interpreted as the predictive density for a future value $\y_{t+1}$ after integrating out the uncertainty about $\unobs$, conditional on the history $\yb^o_{[1:t]}$.

In the SV errors case, Equation~\ref{preddens} is a $(n+p+3)$-dimensional integral which cannot be solved analytically. Nevertheless, it may be evaluated at an arbitrary point $x$ through Monte Carlo integration,
\begin{equation}
 \label{preddensMC}
 p(x|\yb^o_{[1:t]}) \approx \frac{1}{M}\sum_{m=1}^M p(x|\yb^o_{[1:t]}, \unobs^{(m)}_{[1:t]}),
\end{equation}
where $\unobs^{(m)}_{[1:t]}$ stands for the $m$\textsuperscript{th} draw from the respective posterior distribution up to time $t$. If Equation~\ref{preddensMC} is evaluated at $x=y_{t+1}^o$, we refer to it as the (one-step-ahead) \emph{predictive likelihood} at time $t+1$, denoted $\PL_{t+1}$.
Moreover, draws from the posterior predictive distribution can be obtained by simulating values $\y_{t+1}^{(m)}$ from the distribution given through the density $p(\y_{t+1}|\yb^o_{[1:t]}, \unobs^{(m)}_{[1:t]})$, the summands of Equation~\ref{preddensMC}.

For the model at hand, the predictive density and likelihood can thus be computed through the following

\begin{alg}[Predictive density and likelihood evaluation at time $t+1$]
 \label{predlikalg}
 \
 \begin{enumerate}
  \item Reduce the dataset to a training set $\yb^o_{[1:t]} = (\y_1^o,\dots,y_t^o)^\top$.
  \item Run the posterior sampler using data from the training set only to obtain $M$ posterior draws $\unobs^{(m)}_{[1:t]}$, $m=1,\dots, M$.
  \item[(3a.)] Needed for the SV model only: Simulate $M$ values from the conditional distribution $\hsv_{t+1,[1:t]}|\yb^o_{[1:t]},\unobs_{[1:t]}$ by drawing $\hsv_{t+1,[1:t]}^{(m)}$ from a normal distribution with mean $\mupar_{[1:t]}^{(m)} + \phipar_{[1:t]}^{(m)}(\hsv^{(m)}_{t,[1:t]}-\mupar_{[1:t]}^{(m)})$ and standard deviation $\sigmaparextrasub{[1:t]}^{(m)}$ for $m=1,\dots,M$.
  \item[(3b.)] Needed for the GARCH model only: Obtain $M$ draws from the conditional distribution $\sigma_{t+1,[1:t]}|\yb^o_{[1:t]},\unobs_{[1:t]}$ by computing $\sigma_{t+1,[1:t]}^{(m)} = \sqrt{\alpha_{0,[1:t]}^{(m)}+\alpha_{1,[1:t]}^{(m)}\left(\tilde y_t^o\right)^2+\alpha_{2,[1:t]}^{(m)}\left(\sigma_{t,[1:t]}^{(m)}\right)^2}$ for $m=1,\dots,M$. 
  \item[4a.] To obtain $\PL_{t+1}$, average over $M$ densities of normal distributions with mean $(1,\y^o_t)\times\betab^{(m)}_{[1:t]}$ and standard deviation $\exp\{\hsv_{t+1,[1:t]}^{(m)}/2\}$ (SV model), $\sigma_{t+1,[1:t]}^{(m)}$ (GARCH model), or $\sigmaepsextrasub{[1:t]}^{(m)}$ (homoskedastic model), each evaluated at $y^o_{t+1}$, for $m=1,\dots, M$.
  \item[4b.] To obtain $M$ draws from the predictive distribution, simulate from a normal distribution with mean $(1,\y^o_t)\times\betab^{(m)}_{[1:t]}$ and standard deviation $\exp\{\hsv_{t+1,[1:t]}^{(m)}/2\}$ (SV model), $\sigma_{t+1,[1:t]}^{(m)}$ (GARCH model), or $\sigmaepsextrasub{[1:t]}^{(m)}$ (homoskedastic model) for $m=1,\dots,M$.
 \end{enumerate}
\end{alg}

It is worth pointing out that log predictive likelihoods also carry an intrinsic connection to the log \emph{marginal likelihood}, defined through
\begin{eqnarray*}
 \label{ML}
\log \ML = \log p(\yb^o) = \log \int\limits_{\bm K} \!
 p(\yb^o| \unobs) \times
 p(\unobs) 
 \, \dif \unobs.
\end{eqnarray*}
This real number corresponds to the logarithm of the normalizing constant which appears in the denominator of Bayes' law and is often
used for evaluating model evidence. It can straightforwardly be decomposed into the sum of the logarithms of the one-step-ahead predictive likelihoods:
\[
\log \ML = \log p(\yb^o) = \log \prod_{t=1}^n p(\y_t^o|\yb_{[1:t-1]}^o) = \sum_{t=1}^n \log \PL_{t}.
\]
Thus, Algorithm~\ref{predlikalg} provides a conceptually simple way of computing the marginal likelihood. However, these computations are quite costly in terms of CPU time, as they require an individual model fit for each of the $n$ points in time. On the other hand, due to the embarrassingly parallel nature of the task and because of today's comparably easy access to distributed computing environments, this burden becomes manageable. E.g., the computations for the stochastic volatility analysis in this paper have been conducted in less than one hour, using 25 IBM dx360M3 nodes within a cluster of workstations. Implementation in \proglang{R} was achieved through the packages \pkg{parallel} \citep{r:r} and \pkg{snow} \citep{r:sno}.

Cumulative sums of $\log \PL_t$ also allow for model comparison through cumulative log predictive Bayes factors. Letting $\PL_t(A)$ denote the predictive likelihood of model $A$ at time $t$, and $\PL_t(B)$ the corresponding value of model $B$, the cumulative log predictive Bayes factor at time $u$ (and starting point $s$) in favor of model $A$ over model $B$ is straightforwardly given through
\begin{equation}
 \label{clpbf}
\log \left[ \frac{p_A(\yb^o_{[s+1:u]}|\yb^o_{[1:s]})}{p_B(\yb^o_{[s+1:u]}|\yb^o_{[1:s]})} \right] =  \sum_{t=s+1}^u \log \left[ \frac{\PL_t(A)}{\PL_t(B)} \right] =  \sum_{t=s+1}^u [\log \PL_t(A) - \log \PL_t(B)].
\end{equation}

When the cumulative log predictive Bayes factor is positive at a given point in time, there is evidence in favor of model $A$, and vice versa. In this context, information up to time $s$ is regarded as prior information, while out-of-sample predictive evaluation starts at time $s+1$. Note that the usual (overall) log Bayes factor is a special case of Equation~\ref{clpbf} for $s=0$ and $u=n$.

\subsection{Results}

\begin{figure}[!htp]
\begin{center}
 \includegraphics[width=.984\textwidth]{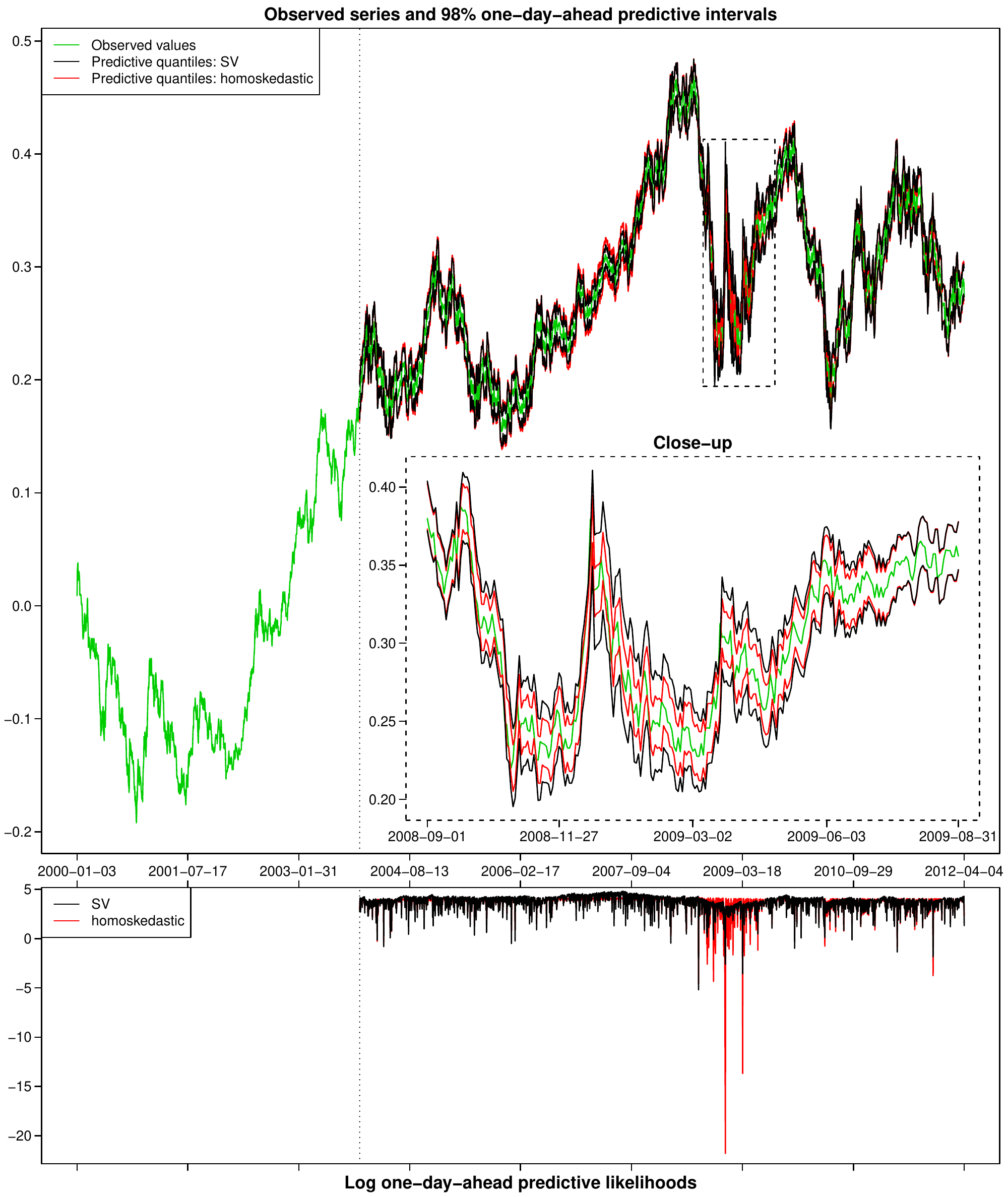}
\end{center}
\caption{Top panel: Observed series (green) and symmetrical 98\% one-day-ahead predictive intervals for the model with homoskedastic errors (red) and the model with SV errors (black). The display also contains a close-up, showing only the period from September 2008 until August 2009. This time span is chosen to include the most noticeable ramifications of the financial crisis. During that phase, predictive performance of the model with homoskedastic errors deteriorates substantially, while SV errors can capture the inflated volatility much better. Bottom panel: Log one-day-ahead predictive likelihoods for both models.}
\label{predlik1}
\end{figure}

In order to reduce prior influence, the first 1000 days are used as a training set only and the evaluation of the predictive distribution starts at $t=1001$, corresponding to December 4, 2003. The homoskedastic model is compared to the model with SV residuals in Figure~\ref{predlik1}. In the top panel, the observed series along with the 98\% one-day-ahead predictive intervals are displayed. The bottom panel shows the log one-day-ahead predictive likelihood. According to these values, SV errors can handle the inflated volatility during that time substantially better. In the course of 2009, the width of the intervals as well as the predictive likelihoods consolidate again. Figure~\ref{predlik1} also contains a close-up of the one-year time span from September 2008 to August 2009. Both credible intervals are similar at the beginning and at the end of this interval. However, there is a substantial difference in early 2009, where SV intervals become around twice as large compared to the corresponding homoskedastic analogs.

A visually barely distinguishable picture emerges when GARCH(1,1) errors are employed instead of SV errors, thus no separate figure is included in this article.

\begin{figure}[!htp]
\begin{center}
 \includegraphics[width=.984\textwidth]{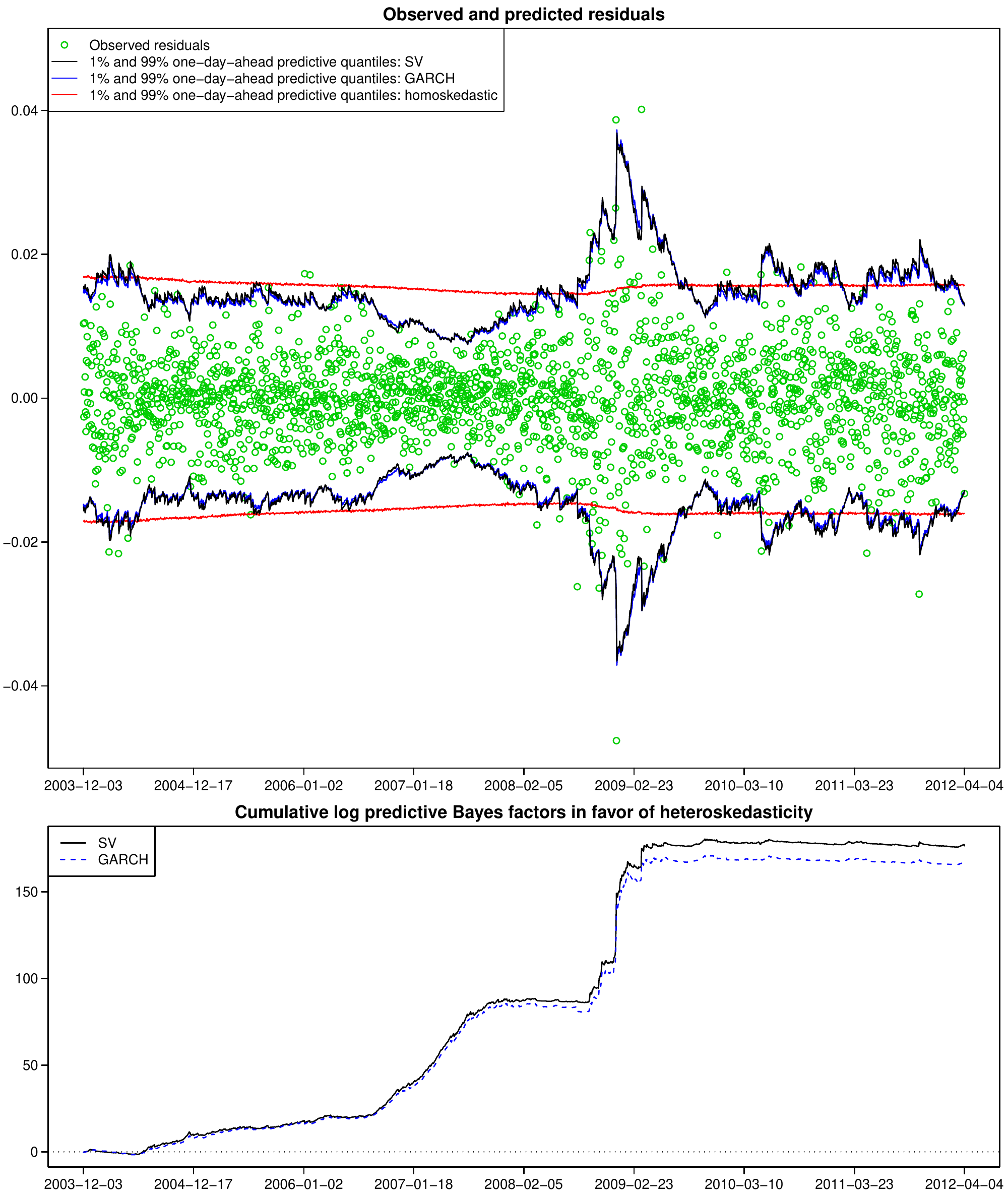}
\end{center}
\caption{Top panel: Observed residuals with respect to the median of the one-day-ahead predictive distribution along with 1\% and 99\% quantiles of the respective predictive distributions. It can clearly be seen that the variances of the predictive distributions in the GARCH and SV models adjust to heteroskedasticity, while the model with homoskedastic errors is much more restrictive. Bottom panel: Cumulative log predictive Bayes factors in favor of the model with SV residuals. Values greater than zero mean that the model with GARCH/SV residuals performs better out of sample up to this point in time.}
\label{predlik3}
\end{figure}

The top panel of Figure~\ref{predlik3} visualizes the performance of the three models; it depicts observed regression residuals against their (one-day-ahead) predicted distributions. For the purpose of this image, observed regression residuals are simply defined as the deviation of the data from the median of the predicted distribution. It stands out that predictive quantiles arising from the models with heteroskedastic errors exhibit much more flexibility to adapt to the ``current state of the world'', while the simpler homoskedastic model barely exhibits this feature.

Generally speaking, there is little difference in predicted residuals until the beginning of 2007. However, during the pre-crisis era (less volatility) and during the actual crisis (more volatility), the models catering for heteroskedasticity perform substantially better. It is interesting to see that predictive intervals for the models with SV errors and GARCH errors are very similar.

In the bottom panel of Figure~\ref{predlik3}, the cumulative sums of the log predictive Bayes factors are displayed. The last points plotted equal to $176.52$ in the SV case and $166.37$ in the GARCH case; this provides overwhelming overall evidence in favor of a model catering for heteroskedasticity and ``decisive'' \citep{jef:the3ed} respectively ``very strong'' \citep{kas-raf:bay} evidence in favor of SV over vanilla GARCH(1,1) with a final cumulative predictive Bayes factor around $25\,000:1$.

It is worth noting that for other exchange rates contained in \code{exrates}, a qualitatively similar picture emerges. For an overview of cumulative log predictive likelihoods, see Table~\ref{exrates}.

\begin{table}[t]
 \centering
 \begin{tabular}{lccc}
  \hline
  \hline
 Currency & SV & GARCH & homoskedastic \\
  \hline
  AUD & 7938 & 7890 & 7554 \\
  CAD & 7851 & 7844 & 7714 \\
  CHF & 9411 & 9337 & 8303 \\
  CZK & 9046 & 8995 & 8659 \\
  DKK\tablefootnote{For exchange rates of the Danish krone with respect to the euro, we analyze \code{1000*log(exrates\$DKK)}, i.e., per mille (\textperthousand) log prices. This way, we avoid convergence issues when obtaining starting values for the MCMC sampler from \pkg{fGarch} that appear otherwise because the krone is pegged very closely to the euro.}
  & 1473 & 1366 & 1178 \\
  GBP & 8568 & 8549 & 8218 \\
  HKD & 7907 & 7897 & 7728 \\
  IDR & 7697 & 7662 & 7269 \\
  JPY & 7607 & 7586 & 7280 \\
  KRW & 7766 & 7749 & 7188 \\
  MXN & 7536 & 7507 & 7055 \\
  MYR & 8082 & 8064 & 7928 \\
  NOK & 8648 & 8616 & 8331 \\
  NZD & 7619 & 7587 & 7440 \\
  PHP & 7890 & 7862 & 7654 \\
  PLN & 8126 & 8094 & 7727 \\
  RON & 9011 & 8880 & 8255 \\
  RUB & 8664 & 8617 & 8146 \\
  SEK & 9110 & 9101 & 8648 \\
  SGD & 8540 & 8529 & 8308 \\
  THB & 7867 & 7844 & 7692 \\
  \hline
  USD & 7878 & 7868 & 7701 \\
  USD [AR(0)]\tablefootnote{These results refer to the analysis of log returns by means of a intercept-only model, i.e., AR(0).}
  & 7876 & 7865 & 7699 \\
  \hline
  \hline
 \end{tabular}
 \caption{Final cumulative predictive log likelihoods for AR(1) models with different error assumptions, applied to the logarithm of several EUR exchange rates. 
 SV is strongly favored in all cases.}
 \label{exrates}
\end{table}

As pointed out above, the analysis of log returns can be viewed upon as a special case of the analysis of log levels where $\beta_1 \equiv 1$ is fixed a priori. When doing so, evidence in favor of heteroskedasticity is again striking. Once more, the model with SV scores highest; its demeaned predictive quantiles are almost indistinguishable from the ``varying $\beta_1$ case''. Predictive quantiles of the model with GARCH residuals again resemble those of the model with SV residuals very closely.
The sum of log predictive likelihoods for $t \in \{1001,1002,\dots,n\}$ is given in the last line of Table~\ref{exrates}. The values are slightly but consistently lower than for the models where $\beta_1$ is estimated from the data. The display of predictive intervals for the log returns is practically identical to Figure~\ref{predlik3} and thus omitted here.

Concluding this somewhat anecdotal prediction exercise, we would like to note that for a thorough and more universal statement concerning the real-world applicability and predictive accuracy of \pkg{stochvol}, further studies with different datasets and a richer variety of competing models, potentially including realized volatility measures, are called for. Such a voluminous undertaking is, however, beyond the scope of this paper.

\section{Conclusion}
The aim of this article was to introduce the reader to the functionality of \pkg{stochvol}. This \proglang{R} package provides a fully Bayesian simulation-based approach for statistical inference in stochastic volatility models. The typical application workflow of \pkg{stochvol} was illustrated through the analysis of exchange rate data contained in the package's \code{exrates} dataset. Furthermore, it was shown how the package can be used as a ``plug-in'' tool for other MCMC samplers. This was illustrated by estimating a Bayesian linear model with SV errors.

In the predictive example, both log levels of exchange rates from EUR to USD and log returns thereof were analyzed. For this dataset, out-of-sample analysis through cumulative predictive Bayes factors clearly showed that modeling regression residuals heteroskedastically substantially improves predictive performance, especially in turbulent times. A direct comparison of SV and vanilla GARCH(1,1) indicated that the former performs better in terms of predictive accuracy.

\section*{Acknowledgments}

The author would like to thank Sylvia Fr\"{u}hwirth-Schnatter, Hedibert Freitas Lopes, Karin Dobernig, and two anonymous referees for helpful comments and suggestions.

\bibliography{mybib}

\begin{thebibliography}{30}
\newcommand{\enquote}[1]{``#1''}
\providecommand{\natexlab}[1]{#1}
\providecommand{\url}[1]{\texttt{#1}}
\providecommand{\urlprefix}{URL }
\expandafter\ifx\csname urlstyle\endcsname\relax
  \providecommand{\doi}[1]{doi:\discretionary{}{}{}#1}\else
  \providecommand{\doi}{doi:\discretionary{}{}{}\begingroup
  \urlstyle{rm}\Url}\fi
\providecommand{\eprint}[2][]{\url{#2}}

\bibitem[{Bollerslev(1986)}]{bol:gen}
Bollerslev T (1986).
\newblock \enquote{Generalized Autoregressive Conditional Heteroskedasticity.}
\newblock \emph{Journal of Econometrics}, \textbf{31}(3), 307--327.
\newblock \doi{10.1016/0304-4076(86)90063-1}.

\bibitem[{Bos(2012)}]{bos:rel}
Bos CS (2012).
\newblock \enquote{Relating Stochastic Volatility Estimation Methods.}
\newblock In L~Bauwens, C~Hafner, S~Laurent (eds.), \emph{Handbook of
  Volatility Models and Their Applications}, pp. 147--174. John Wiley \& Sons.
\newblock \doi{10.1002/9781118272039.ch6}.

\bibitem[{Eddelbuettel and Fran\c{c}ois(2011)}]{edd-fra:rcp}
Eddelbuettel D, Fran\c{c}ois R (2011).
\newblock \enquote{{\pkg{Rcpp}}: Seamless {\proglang{R}} and {\proglang{C++}}
  Integration.}
\newblock \emph{Journal of Statistical Software}, \textbf{40}(8), 1--18.
\newblock \doi{10.18637/jss.v040.i08}.

\bibitem[{Engle(1982)}]{eng:aut}
Engle RF (1982).
\newblock \enquote{Autoregressive Conditional Heteroscedasticity With Estimates
  of the Variance of {U}nited {K}ingdom Inflation.}
\newblock \emph{Econometrica}, \textbf{50}(4), 987--1007.
\newblock \doi{10.2307/1912773}.

\bibitem[{Fr{\"u}hwirth-Schnatter and Wagner(2010)}]{fru-wag:sto}
Fr{\"u}hwirth-Schnatter S, Wagner H (2010).
\newblock \enquote{Stochastic Model Specification Search for {G}aussian and
  Partial Non-{G}aussian State Space Models.}
\newblock \emph{Journal of Econometrics}, \textbf{154}(1), 85--100.
\newblock \doi{10.1016/j.jeconom.2009.07.003}.

\bibitem[{Genz \emph{et~al.}(2013)Genz, Bretz, Miwa, Mi, Leisch, Scheipl, and
  Hothorn}]{r:mvt}
Genz A, Bretz F, Miwa T, Mi X, Leisch F, Scheipl F, Hothorn T (2013).
\newblock \emph{\pkg{mvtnorm}: Multivariate Normal and t Distributions}.
\newblock \proglang{R} package version 0.9-9996,
  \urlprefix\url{http://CRAN.R-project.org/package=mvtnorm}.

\bibitem[{Geweke and Amisano(2010)}]{gew-ami:com}
Geweke J, Amisano G (2010).
\newblock \enquote{Comparing and Evaluating {B}ayesian Predictive Distributions
  of Asset Returns.}
\newblock \emph{International Journal of Forecasting}, \textbf{26}(2),
  216--230.
\newblock \doi{10.1016/j.ijforecast.2009.10.007}.

\bibitem[{Ghysels \emph{et~al.}(1996)Ghysels, Harvey, and
  Renault}]{ghy-etal:sto}
Ghysels E, Harvey AC, Renault E (1996).
\newblock \enquote{Stochastic Volatility.}
\newblock In GS~Maddala, CR~Rao (eds.), \emph{Statistical Methods in Finance},
  volume~14 of \emph{Handbook of Statistics}, pp. 119--191. Elsevier.
\newblock \doi{10.1016/S0169-7161(96)14007-4}.

\bibitem[{Hills and Smith(1992)}]{hil-smi:par}
Hills SE, Smith AFM (1992).
\newblock \enquote{Parameterization Issues in {B}ayesian Inference.}
\newblock In JM~Bernardo, JO~Berger, AP~Dawid, AFM Smith (eds.),
  \emph{{P}roceedings of the {F}ourth {V}alencia {I}nternational {M}eeting},
  volume~4 of \emph{Bayesian {S}tatistics}, pp. 227--246. Oxford University
  Press.

\bibitem[{Jacquier \emph{et~al.}(1994)Jacquier, Polson, and
  Rossi}]{jac-etal:bayJBES}
Jacquier E, Polson NG, Rossi PE (1994).
\newblock \enquote{Bayesian Analysis of Stochastic Volatility Models.}
\newblock \emph{Journal of Business {\rm \&} Economic Statistics},
  \textbf{12}(4), 371--389.
\newblock \doi{10.1080/07350015.1994.10524553}.

\bibitem[{Jacquier \emph{et~al.}(2004)Jacquier, Polson, and
  Rossi}]{jac-etal:bayJE}
Jacquier E, Polson NG, Rossi PE (2004).
\newblock \enquote{Bayesian Analysis of Stochastic Volatility Models With
  Fat-Tails and Correlated Errors.}
\newblock \emph{Journal of Econometrics}, \textbf{122}(1), 185--212.
\newblock \doi{10.1016/j.jeconom.2003.09.001}.

\bibitem[{Jeffreys(1961)}]{jef:the3ed}
Jeffreys H (1961).
\newblock \emph{Theory of Probability}.
\newblock Third edition. Oxford University Press.

\bibitem[{Kass and Raftery(1995)}]{kas-raf:bay}
Kass RE, Raftery AE (1995).
\newblock \enquote{{B}ayes Factors.}
\newblock \emph{Journal of the American Statistical Association}, \textbf{90},
  773--795.
\newblock \doi{10.1080/01621459.1995.10476572}.

\bibitem[{Kastner(2016{\natexlab{a}})}]{kas:dea}
Kastner G (2016{\natexlab{a}}).
\newblock \enquote{Dealing with Stochastic Volatility in Time Series Using the
  \proglang{R} Package \pkg{stochvol}.}
\newblock \emph{Journal of Statistical Software}, \textbf{69}(5), 1--30.
\newblock \doi{10.18637/jss.v069.i05}.

\bibitem[{Kastner(2016{\natexlab{b}})}]{r:sto}
Kastner G (2016{\natexlab{b}}).
\newblock \emph{\pkg{stochvol}: Efficient Bayesian Inference for Stochastic
  Volatility (SV) Models}.
\newblock \proglang{R} package version 1.2.3,
  \urlprefix\url{http://CRAN.R-project.org/package=stochvol}.

\bibitem[{Kastner and Fr{\"u}hwirth-Schnatter(2014)}]{kas-fru:anc}
Kastner G, Fr{\"u}hwirth-Schnatter S (2014).
\newblock \enquote{{A}ncillarity-Sufficiency Interweaving Strategy ({ASIS}) for
  Boosting {MCMC} Estimation of Stochastic Volatility Models.}
\newblock \emph{Computational Statistics \& Data Analysis}, \textbf{76},
  408--423.
\newblock \doi{10.1016/j.csda.2013.01.002}.

\bibitem[{Kastner \emph{et~al.}(2014)Kastner, Fr\"{u}hwirth-Schnatter, and
  Lopes}]{kas-etal:ana}
Kastner G, Fr\"{u}hwirth-Schnatter S, Lopes HF (2014).
\newblock \enquote{Analysis of Exchange Rates via Multivariate {B}ayesian
  Factor Stochastic Volatility Models.}
\newblock In E~Lanzarone, F~Ieva (eds.), \emph{The Contribution of Young
  Researchers to {B}ayesian Statistics -- Proceedings of {BAYSM2013}},
  volume~63 of \emph{Springer Proceedings in Mathematics \& Statistics}, pp.
  181--185. Springer-Verlag.
\newblock \doi{10.1007/978-3-319-02084-6_35}.

\bibitem[{Kim \emph{et~al.}(1998)Kim, Shephard, and Chib}]{kim-etal:sto}
Kim S, Shephard N, Chib S (1998).
\newblock \enquote{Stochastic Volatility: Likelihood Inference and Comparison
  With {ARCH} Models.}
\newblock \emph{Review of Economic Studies}, \textbf{65}(3), 361--393.
\newblock \doi{10.1111/1467-937X.00050}.

\bibitem[{Markowitz(1952)}]{mar:por}
Markowitz H (1952).
\newblock \enquote{Portfolio Selection.}
\newblock \emph{The Journal of Finance}, \textbf{7}(1), 77--91.
\newblock \doi{10.1111/j.1540-6261.1952.tb01525.x}.

\bibitem[{{McC}ausland \emph{et~al.}(2011){McC}ausland, Miller, and
  Pelletier}]{mcc-etal:sim}
{McC}ausland WJ, Miller S, Pelletier D (2011).
\newblock \enquote{Simulation Smoothing for State-Space Models: {A}
  Computational Efficiency Analysis.}
\newblock \emph{Computational Statistics and Data Analysis}, \textbf{55}(1),
  199--212.
\newblock \doi{10.1016/j.csda.2010.07.009}.

\bibitem[{Meyer and Yu(2000)}]{mey-yu:bug}
Meyer R, Yu J (2000).
\newblock \enquote{{\proglang{BUGS}} for a {B}ayesian Analysis of Stochastic
  Volatility Models.}
\newblock \emph{The Econometrics Journal}, \textbf{3}(2), 198--215.
\newblock \doi{10.1111/1368-423X.00046}.

\bibitem[{Omori \emph{et~al.}(2007)Omori, Chib, Shephard, and
  Nakajima}]{omo-etal:sto}
Omori Y, Chib S, Shephard N, Nakajima J (2007).
\newblock \enquote{Stochastic Volatility With Leverage: {F}ast and Efficient
  Likelihood Inference.}
\newblock \emph{Journal of Econometrics}, \textbf{140}(2), 425--449.
\newblock \doi{10.1016/j.jeconom.2006.07.008}.

\bibitem[{Plummer \emph{et~al.}(2006)Plummer, Best, Cowles, and
  Vines}]{plu-etal:cod}
Plummer M, Best N, Cowles K, Vines K (2006).
\newblock \enquote{{\pkg{coda}}: Convergence Diagnosis and Output Analysis for
  {MCMC}.}
\newblock \emph{\proglang{R} News}, \textbf{6}(1), 7--11.
\newblock \urlprefix\url{http://CRAN.R-project.org/doc/Rnews/Rnews_2006-1.pdf}.

\bibitem[{{\proglang{R} Core Team}(2016)}]{r:r}
{\proglang{R} Core Team} (2016).
\newblock \emph{\proglang{R}: A Language and Environment for Statistical
  Computing}.
\newblock \proglang{R} Foundation for Statistical Computing, Vienna, Austria.
\newblock \urlprefix\url{http://www.R-project.org/}.

\bibitem[{Rue(2001)}]{rue:fas}
Rue H (2001).
\newblock \enquote{Fast Sampling of {G}aussian {M}arkov Random Fields.}
\newblock \emph{Journal of the Royal Statistical Society B}, \textbf{63}(2),
  325--338.
\newblock \doi{10.1111/1467-9868.00288}.

\bibitem[{Taylor(1982)}]{tay:fin}
Taylor SJ (1982).
\newblock \enquote{Financial Returns Modelled by the Product of Two Stochastic
  Processes: A Study of Daily Sugar Prices 1691--79.}
\newblock In OD~Anderson (ed.), \emph{Time Series Analysis: Theory and Practice
  1}, pp. 203--226. North-Holland, Amsterdam.

\bibitem[{Tierney \emph{et~al.}(2013)Tierney, Rossini, Li, and
  Sevcikova}]{r:sno}
Tierney L, Rossini AJ, Li N, Sevcikova H (2013).
\newblock \emph{\pkg{snow}: Simple Network of Workstations}.
\newblock \proglang{R} package version 0.3-12,
  \urlprefix\url{http://CRAN.R-project.org/package=snow}.

\bibitem[{Wuertz \emph{et~al.}(2013)Wuertz, Chalabi, Miklovic, Boudt, Chausse,
  and {others}}]{r:fGarch}
Wuertz D, Chalabi Y, Miklovic M, Boudt C, Chausse P, {others} (2013).
\newblock \emph{\pkg{fGarch}: Rmetrics - Autoregressive Conditional
  Heteroskedastic Modelling}.
\newblock \proglang{R} package version 3010.82,
  \urlprefix\url{http://CRAN.R-project.org/package=fGarch}.

\bibitem[{Yu(2005)}]{yu:on}
Yu J (2005).
\newblock \enquote{On Leverage in a Stochastic Volatility Model.}
\newblock \emph{Journal of Econometrics}, \textbf{127}(2), 165--178.
\newblock \doi{10.1016/j.jeconom.2004.08.002}.

\bibitem[{Yu and Meng(2011)}]{yu-men:cen}
Yu Y, Meng XL (2011).
\newblock \enquote{To Center or Not to Center: That is Not the Question---An
  Ancillarity-Suffiency Interweaving Strategy {(ASIS)} for Boosting {MCMC}
  Efficiency.}
\newblock \emph{Journal of Computational and Graphical Statistics},
  \textbf{20}(3), 531--570.
\newblock \doi{10.1198/jcgs.2011.203main}.

\end{thebibliography}
\end{document}